\documentclass[sigconf]{acmart}
\AtBeginDocument{%
  \providecommand\BibTeX{{%
    \normalfont B\kern-0.5em{\scshape i\kern-0.25em b}\kern-0.8em\TeX}}}

\copyrightyear{2021}
\acmYear{2021}
\setcopyright{rightsretained}
\acmConference[MobiHoc '21]{The Twenty-second International Symposium on Theory, Algorithmic Foundations, and Protocol Design for Mobile Networks and Mobile Computing}{July 26--29, 2021}{Shanghai, China}
\acmBooktitle{The Twenty-second International Symposium on Theory, Algorithmic Foundations, and Protocol Design for Mobile Networks and Mobile Computing (MobiHoc '21), July 26--29, 2021, Shanghai, China}\acmDOI{10.1145/3466772.3467052}
\acmISBN{978-1-4503-8558-9/21/07}

\usepackage{algorithm}
\usepackage{graphicx}
\usepackage{amsmath}
\usepackage{amsthm}
\usepackage{algorithmic}
\usepackage{array}
\usepackage{subcaption}
\usepackage{gensymb}
\usepackage{url}
\usepackage{multicol}
\usepackage{multirow}
\usepackage{float}
\usepackage{framed}
\usepackage{xcolor}
\usepackage{mathtools}
\usepackage{etoolbox}

\newtheorem{proposition}{Proposition}[section]

\newrobustcmd{\bb}{\bfseries}

\begin{document}

\title{Man-in-the-Middle Attack Resistant Secret Key Generation via Channel Randomization}

\author{Yanjun Pan}
\affiliation{%
  \institution{University of Arizona
  }
  \city{Tucson}
  \state{AZ}
  \country{USA}
}
\email{yanjunpan@email.arizona.edu}

\author{Ziqi Xu}
\affiliation{%
  \institution{University of Arizona}
  \city{Tucson}
  \state{AZ}
  \country{USA}
}
\email{ zxu1969@email.arizona.edu}

\author{Ming Li}
\affiliation{%
  \institution{University of Arizona}
  \city{Tucson}
  \state{AZ}
  \country{USA}}
\email{lim@arizona.edu}

\author{Loukas Lazos}
\affiliation{%
 \institution{University of Arizona}
 \city{Tucson}
  \state{AZ}
  \country{USA}}
\email{llazos@arizona.edu}

\renewcommand{\shortauthors}{Yanjun Pan, Ziqi Xu, Ming Li, and Loukas Lazos}

\begin{abstract}
Physical-layer based key generation schemes  exploit the channel reciprocity for secret key extraction,  which can achieve information-theoretic secrecy against eavesdroppers. Such methods, although practical, have been shown to be vulnerable against man-in-the-middle (MitM) attacks, where an active adversary, Mallory, can influence and infer part of the secret key generated between Alice and Bob by injecting her own packet upon observing highly correlated channel/RSS measurements from  Alice and Bob. As all the channels remain stable within the channel coherence time, Mallory's injected packets cause Alice and Bob to measure similar RSS, which allows Mallory to successfully predict the derived key bits. To defend against such a MitM attack, we propose to utilize a reconfigurable antenna at one of the legitimate transceivers to proactively randomize the channel state across different channel probing rounds. The randomization of the antenna mode at every probing round breaks the temporal correlation of the channels from the adversary to the legitimate devices, while preserving the reciprocity of the channel between the latter. This prevents key injection from the adversary without affecting Alice and Bob's ability to measure common randomness.  We theoretically analyze the security of the protocol and conduct extensive simulations and real-world experiments to evaluate its performance. Our results show that our approach eliminates the advantage of an active MitM attack by driving down the probability of successfully guessing bits of the secret key to a random guess.
\end{abstract}





\maketitle

\section{Introduction}
In the past two decades, the idea of physical-layer  key generation has drawn significant attention from the security community \cite{mathur2008radio,jana2009effectiveness,zeng2010exploiting,wang2011fast}. To generate a pairwise key, Alice and Bob measure the inherent common randomness in the physical wireless channel. This randomness is unique to the location and time that measurements take place. Physical-layer  key generation has been  shown to be information-theoretically secure against passive eavesdroppers and does  not assume any prior shared secrets between Alice and Bob, nor does it require a public key infrastructure. It complements upper-layer security primitives and has been touted as a less expensive and more flexible solution to replace public-key cryptography than competing alternatives (e.g., quantum cryptography \cite{mathur2008radio}). 

The key assumptions  exploited by  physical-layer key generation schemes are   channel reciprocity and spatial signal decorrelation. Specifically, the channel reciprocity property implies that the signal distortion (attenuation, delay, phase shift, and fading) is identical in both directions of a link, which enables the measurement of common randomness by legitimate devices. The spatial decorrelation property indicates that in rich scattering environments,  two receivers located a few wavelengths away will experience  uncorrelated  channels. These properties have been demonstrated to hold in practice and the latter is essential for securing the physical-layer key generation process against eavesdroppers \cite{marino2014secret, zenger2015security}. However,  an active attacker can bypass this assumption by injecting its own signals to influence and infer the derived key bits between Alice and Bob. For example, Eberz {\em et al.}   \cite{eberz2012practical} proposed a practical man-in-the-middle (MitM) attack against RSS-based key generation protocols \cite{mathur2008radio,jana2009effectiveness}, where Mallory exploits the same channel characteristics as Alice and Bob do. Mallory awaits for attack opportunities when the channel
from Mallory-to-Alice and Mallory-to-Bob are similar, to inject packets that cause a similar channel measurement at both Alice and Bob. This simple and practical attack strategy enables the adversary to recover
up to 47\% of the secret key bits generated.

The root cause of this vulnerability is that even in relatively rich scattering environments, the coherence time of the channel is likely to span multiple channel probing rounds. This allows the adversary to measure for opportunities over one probing round before attacking the subsequent probing round with high probability of predicting the extracted bits due to the channel similarity. In this paper, we aim at defending against such MitM attacks, by proposing a channel randomization-based key generation protocol. 
With the key observation that the MitM attack has to be launched in a \textit{wait-then-attack} manner, our idea is to break the channel coherence and reciprocity across different probing rounds to prevent successful bit inference/injection from a MitM attacker, while preserving these properties within each probing round for key generation. To realize this goal, we utilize a reconfigurable antenna at one of the legitimate transceivers, where in each probing round the antenna mode is randomly chosen from a large set of diverse antenna modes, which effectively randomizes the channel to/from the adversary.

The main contributions of this work are three-fold:
\begin{itemize}
   \item We are the first to propose a channel randomization-based  approach  to prevent active MitM attacks against physical layer key generation protocols. The main innovation is in the quantization phase of the protocol, where one of the legitimate devices is equipped with  a pattern reconfigurable antenna. Randomly switching the antenna mode randomizes the channel state in each probing round, thus preventing the adversary from predicting the outcome of packet injection attacks and consequently preventing the adversary from predicting bits of the secret key. 
   
   \item We theoretically analyze the security of our scheme against the MitM attack by deriving key security metrics such as the key efficiency rate, the key recovery rate, and the probability of correctly guessing the entire key generated by Alice and Bob, under realistic multipath channel models. 
   
   \item We  validate  our theoretical analysis using extensive simulation and real-world experiments with commodity devices. Empirical results show that our channel randomization-based key generation approach greatly reduces the successful key bit guessing probability of the MitM attack, making it less effective than random guess.
\end{itemize}


This  paper  is  organized  as  follows: In Sec. 2, we give an overview of existing physical-layer key generation protocols and the practical MitM attack proposed in \cite{eberz2012practical}. We present our channel randomization based physical-layer key generation protocol in Sec. 3. We theoretically analyze its security in Sec. 4. In Sec. 5, we evaluate the protocol security experimentally. In Sec. 6, we summarize the related works. We present a discussion on a more advanced adversary model  in Sec. 6 and conclude in Sec. 7.

\section{Preliminaries}
\label{sec:protocol}

Signals traveling over a wireless channel are modified by the channel in a way that is unique to the transmit-receive pair. Physical-layer based key generation protocols exploit this radio propagation characteristic to privately extract secret bits between two parties, Alice and Bob, who do not share any prior secrets. The majority of physical-layer key generation protocols consist of the three basic phases shown in Fig.~\ref{fig:diagram1}, namely the quantization phase, the information reconciliation phase, and the key verification phase. We describe each phase in detail.

\subsection{Physical-layer Key Generation} 

\subsubsection{Quantization Phase} 
The quantization phase aims at converting  channel measurements between Alice and Bob to an initial bit stream that will be used for key generation. This is a two-step process involving probing and quantization. During probing,  Alice and Bob  exchange probing signals to measure channel properties such as received signal strength (RSS) or channel state information (CSI).

\begin{figure}[t]
\vspace{-0.1in}
\begin{center}
\includegraphics[width=0.45\textwidth]{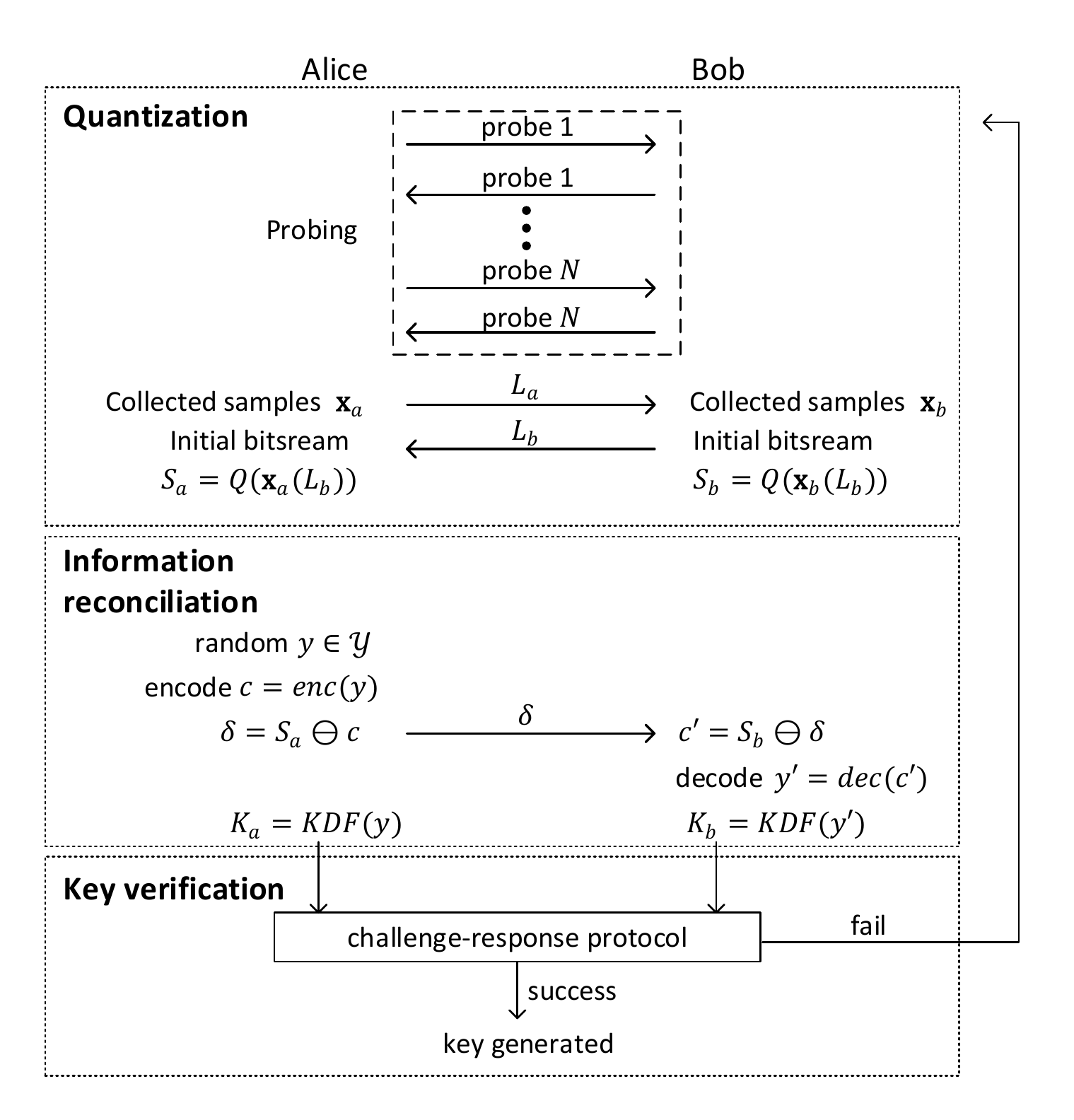}
\end{center}
\vspace{-15pt}
\caption{Three typical phases for existing physical-layer based key generation protocols}
\label{fig:diagram1}
\vspace{-15pt}
\end{figure}

In practice, Alice transmits a probe first, allowing Bob to obtain a channel measurement.  The probing signal contains a preamble that enables  synchronization, channel estimation, and frequency offset correction at Bob, as well as the ID of Alice  for sender identification  (although no sender authentication is provided). Other payload fields may be included. Bob samples the received signal, which can be represented as
\begin{align}
    r_b^a(t_i) & = h_{ba}(t_i)s(t_i) + n_b(t_i) \label{eq:rss1}
\end{align}
where $r_b^a(t_i)$ denotes the signal received at Bob  during time $t_i$  when Alice transmits, $h_{ba}(t_i)$ is the impulse response of the $A-B$ channel at time $t_i$, $s(t_i)$ is the known probe signal transmitted by Alice, and $n_b(t_i)$ is the noise at Bob. From the received signal samples, Bob can further extract the relevant channel measurements that will be used for quantization such as the average RSS or the CSI amplitude. We denote a channel measurement obtained by Bob over a probe packet during the $i^{th}$ round as $x_b(i).$ Similarly, Bob transmits its one probe to Alice, allowing her to obtain channel measurement $x_a(i)$. The exchange of two probes completes one {\em probing round.}
After $N$ probing rounds, Alice and Bob have extracted $N$ channel measurements represented by vectors $\mathbf{x}_a = \{x_a(1), x_a(2), \dots, x_a(N) \}$ and $\mathbf{x}_b = \{x_b(1), x_b(2), \dots, x_b(N) \}$,  respectively.

With the completion of the $N^{th}$ probing round, Alice and Bob proceed to quantization by independently selecting quantization thresholds. A commonly-used method is the multi-threshold approach  \cite{mathur2008radio,eberz2012practical}, where a device $u$ independently selects two thresholds $q_+^u$ and $q_-^u$ as:
\begin{align}
& q_+^u = \mu(\mathbf{x}_u) + \beta\sigma(\mathbf{x}_u) & q_-^u = \mu(\mathbf{x}_u) - \beta\sigma(\mathbf{x}_u).
\label{eq:q}
\end{align}
Here, $\mu(\mathbf{x}_u)$ and $\sigma(\mathbf{x}_u)$ are the mean and standard deviation of $\mathbf{x}_u$, respectively and $\beta$ is a weight parameter with $0 < \beta < 1$. The quantization function $Q(x)$ for a measurement $x$ is defined as 
\begin{align}
& Q(x) =\left\{
    \begin{aligned}
        0, &\quad x < q_- \\
        1, &\quad x > q_+\\
    \end{aligned}
\right.
\label{quant}
\end{align}
To improve robustness, the quantization can be implemented with \textit{excursions} \cite{mathur2008radio,jana2009effectiveness}, which are defined as  $e$
consecutive channel measurements in $\mathbf{x}_a$ (or $\mathbf{x}_b$) that exceed $q_+$ or are below $q_-$. To identify excursions,  
Alice first goes through $\mathbf{x}_a$, determines
the locations of excursions, and publicly sends the index list of excursion locations $L_a$ to Bob. Bob checks his own measurements  $\mathbf{x}_b$ at the indices specified in
$L_a$ to determine whether an excursion also occurs. Bob then identifies all indexes in $L_a$
that also produce an excursion in $\mathbf{x}_b$, and sends these indexes to Alice in list $L_b$. Finally, Alice and Bob quantize each excursion in $L_b$ and construct the initial bitstreams $S_a$ and $S_b$, respectively.

\subsubsection{Information Reconciliation Phase}

Ideally, if the channel between Alice and Bob is perfectly reciprocal, the quantized sequences $S_a$ and $S_b$ at Alice and Bob would match perfectly. In practice, there is a mismatch between $S_a$ and $S_b$ because the channel is measured at different times at Alice and Bob (probes are sequentially transmitted) and also due to hardware differences. This results to a small number of mismatched bits in the initial  bitstreams. To correct these bit mismatches and agree on a common key, Alice and Bob implement an information reconciliation phase.
Many information reconciliation protocols have been proposed in the literature. In Fig. \ref{fig:diagram1}, we show one adopted by several schemes, proposed by Sch\"urmann
and Sigg in \cite{schurmann2011secure}.  The scheme is based on the fuzzy vault construction \cite{juels2006fuzzy}. It  Reed-Solomon codes to reconcile two sequences that differ in a number of bits below the error-correcting capacity. Further details can be found in Appendix, Sec. 8.1.

\subsubsection{Key Verification Phase}
Finally, both parties use a simple challenge-response
protocol to verify that the established secret keys are identical. Unsuccessful verification results in a key disagreement and the protocol is repeated.

\subsection{Adversary Model} \label{sec:MitM}
{\bf Adversary goal:} In this paper, we focus on an adversary whose goal is to infer the bits in the pairwise key agreed between Alice and Bob. By learning the symmetric key, the adversary can later compromise the secrecy of the communications between the two parties. The bit inference can be achieved either through passive eavesdropping \cite{edman2011passive,steinmetzer2015lockpicking}  or active attacks \cite{eberz2012practical,zhou2014secret}. Note, that we do not address other type of adversary goals such as authentication. This security property can be achieved in conjunction with key generation using a number of additional out-of-band or in-band methods \cite{mccune2005seeing,cagalj2006integrity,17TDSCYanjun}.  

{\bf Passive attacks:} In a passive attack, one or more  eavesdroppers overhear the  probes of Alice and Bob to infer the generated secret key, without actively interrupting/modifying their communication. This attack model has been widely studied in past work (e.g., \cite{edman2011passive,steinmetzer2015lockpicking}). The main defense against eavesdropping  exploits the fast channel decorrelation with distance in rich scattering environments. When Mallory is situated several wavelengths away from Alice and Bob, the channel becomes independent from that between the legitimate parties, thus preventing the correct channel measurement and quantization.  

\begin{figure}[t]
\vspace{-0.1in}
\begin{center}
\includegraphics[width=0.5\textwidth]{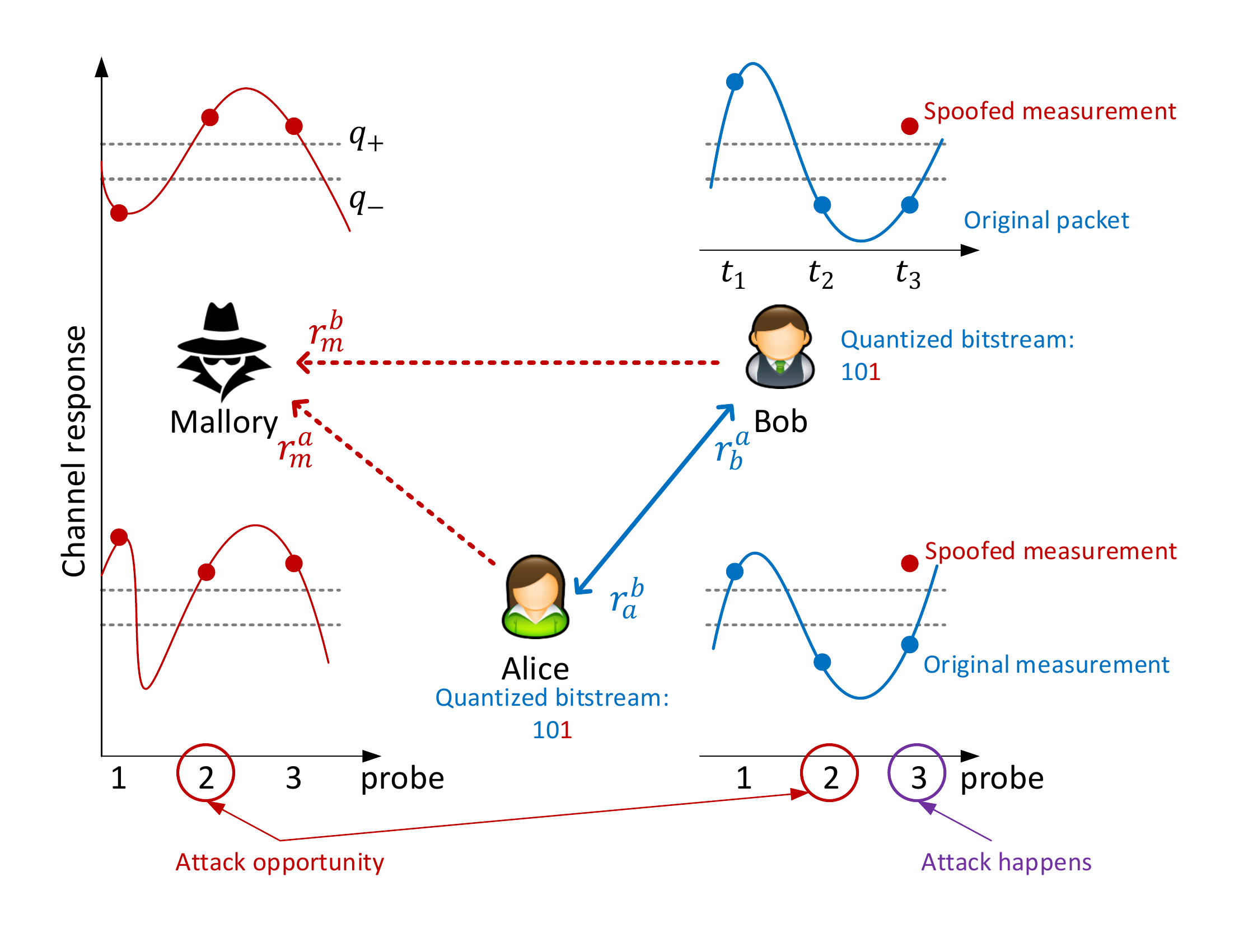}
\end{center}
\vspace{-0.2in}
\caption{Mallory identifies an attack opportunity at round 2 by measuring a similar channel from Alice and Bob. In the follow-up probing rounds, Mallory injects her own probes to influence the measurements at Alice and Bob. Alice and Bob convert the measurement at $t_3$ to bit 1, which can be predicted by Mallory.}
\label{fig:system}
\vspace{-0.2in}
\end{figure}

{\bf Active attacks:} In this paper, we  focus on  active MitM attacks that have been shown to be successful even if the adversary is several wavelengths away from Alice and Bob.  Unlike traditional MitM attacks, the attacker does not aim at establishing a separate key with Alice and Bob, since that would require  blocking any direct communication between Alice and Bob. Moreover, such a MitM attack are easily detected when authentication is incorporated into key generation.  For instance, if authentication is performed by a visual out-of-band channel, the establishment of different key between Alice and Bob is detected.  

A practical MitM attack for key inference is proposed by Eberz {\em et al.} \cite{eberz2012practical}. In this attack, Mallory first identifies an attack opportunity on probing round $i$. She then reactive jams the transmitted probes on  round $i+1$ and injects her own probes. This is possible because probes cannot be authenticated in the absence of a common secret. Specifically,  let $\text{RSS}_x^y$ denote the RSS measured at $x$ when $y$ transmits. Mallory identifies an attack opportunity during the $i^{th}$ probing round when the following condition is met \cite{eberz2012practical}:
\begin{equation}
\left( |\text{RSS}_m^a - \text{RSS}_m^b| < d \right) ~\textbf{AND}~ \left(\text{RSS}_m^a > q_+ ~\textbf{OR}~ \text{RSS}_m^a < q_- \right) 
\label{eq:opty}
\end{equation}
where $d$ is a threshold that defines the maximum RSS difference between probes. The opportunity incorporates two criteria. 
The first criterion requires the difference between RSS values received from Alice and Bob to be smaller than $d$, which indicates a similar channel from Mallory to both Alice and Bob. 
The second criterion detects if the RSS measurements exceed any of the quantization thresholds, so that Alice and Bob are likely to use that round to extract a bit during quantization. 
This criterion also decides the bit that is guessed by Mallory. If the injection attack is implemented after observing $\text{RSS}_m^a > q_+$, then Mallory guesses the injected bit at round $i+1$ to be 1, otherwise Mallory guesses the injected bit to be  0.  {\em The prerequisite for successfully guessing  the bit extracted by Alice and Bob during round $i+1$ is that the channel remains the same as in round $i$.} If both rounds are within the channel coherence time, Mallory can predict the RSS at both Alice and Bob and infer the extracted bit with high probability. 

Figure \ref{fig:system} shows an example of this attack strategy, where an attack opportunity is found at round 2, with $ \text{RSS}_m^a(2) \approx \text{RSS}_m^b(2)$ and $ \text{RSS}_m^a(2) > q_+$. Mallory follows-up with a jamming and injection attack  in the next round. 
Due to the channel coherence, it follows that $\text{RSS}_m^a(3) \approx \text{RSS}_m^a(2)$ and $\text{RSS}_m^b(3) \approx \text{RSS}_m^b(2)$. Besides, $\text{RSS}_a^m(3) \approx \text{RSS}_m^a(2)$ and $\text{RSS}_b^m(3) \approx \text{RSS}_m^b(3)$ due to channel reciprocity. Hence, at round 3, the RSS values at Alice and Bob are $\text{RSS}_a^m(3) \approx \text{RSS}_m^a(2) > q_+$ and $\text{RSS}_b^m(3) \approx \text{RSS}_a^m(3) > q_+$ since $ \text{RSS}_m^a(2) \approx \text{RSS}_m^b(2)$. Correspondingly, the  measurement at round 3 will be converted into bit 1, and Mallory can correctly guess it.  Note that although the MitM proposed in \cite{eberz2012practical} is implemented based on RSS, it can be easily extend to CSI by letting Mallory find opportunities upon the amplitude of CSI.

\section{RAKG: RA-based Key Generation} \label{sec:our_protocol}

The main vulnerability exploited by the MitM attack described in the previous section stems from the similarity of the M-A and M-B channels over several probing rounds, once an opportunity is identified. This allows Mallory to predict the impact of the injected probes with high probability. To defend against this attack, our basic idea is to reduce the channel coherence time to the duration of a single probing round by using a reconfigurable antenna (RA) at either Alice or Bob. Without loss of generality, let Alice being equipped with an RA with a total of $U$ antenna modes. By randomly selecting the antenna mode $u$ on every probing round, Alice can randomize the M-A (and A-B) channel on every probing round. Therefore, even if Mallory identifies an opportunity on the $i^{th}$ round, the channel has changed when she performs the injection attack on round $i+1$. 
On the other hand, the A-B channel  remains reciprocal within the same   probing round, so that Alice and Bob  can still extract a common bit from the randomized channel.

\subsection{Channel Randomization with an RA}
An RA is one antenna that can swiftly reconfigure its radiation  pattern, polarization, and frequency, or combinations of them by rearranging its antenna currents \cite{bernhard2007reconfigurable}. We choose an RA for channel randomization as it provides a  diversity of antenna states \cite{costantine2015reconfigurable}.
For example, a type of parasitic-layer based pattern-reconfigurable antenna uses p-i-n diodes as switches, where 12 switches can give up to $2^{12} = 4096$ different configurations  \cite{costantine2015reconfigurable}. 

Let Alice be equipped with an RA, whereas  Bob is equipped with a conventional omnidirectional antenna (OA). Mallory can have any type  of antenna, including OA, directional antenna, or an RA.   The time-varying multipath channel that incorporates the impact of an RA  is described as \cite{lusina2009antenna}:
\begin{align}
    h(u_j, t_i) & = \sum\limits_{l=0}^{P}g(u_j,\theta_l)a_l(t_i), \label{eq:h_u_t}
\end{align}
where $u_j$ is the antenna mode selected,  $P$ is the  number of signal paths, $g(u_j,\theta_l)$ is the antenna gain under mode $u_j$ at direction $\theta_l$, and $a_l(t_i)$ is the fading parameter of the $l$-th path at time $t_i$. More specifically, $a_l(t_i) = \alpha_{l}(t_i)e^{-j\phi_{l}(t_i)}$,  where $\alpha_{l}(t_i)$ and $\phi_{l}(t_i)$ are the amplitude and phase of the fading, respectively.
From Eq. \eqref{eq:h_u_t}, we can see that although Mallory can know $g(u_j,\theta_l)$ for any given antenna mode,  she does not know which antenna mode is selected by Alice. Therefore, the A-B and M-A channels will be randomized between probing rounds. However, the M-B channel will remain the same between several probing rounds because  Bob is equipped with an OA. 
\begin{figure}[t]
\vspace{-0.15in}
\begin{center}
\includegraphics[width=0.45\textwidth]{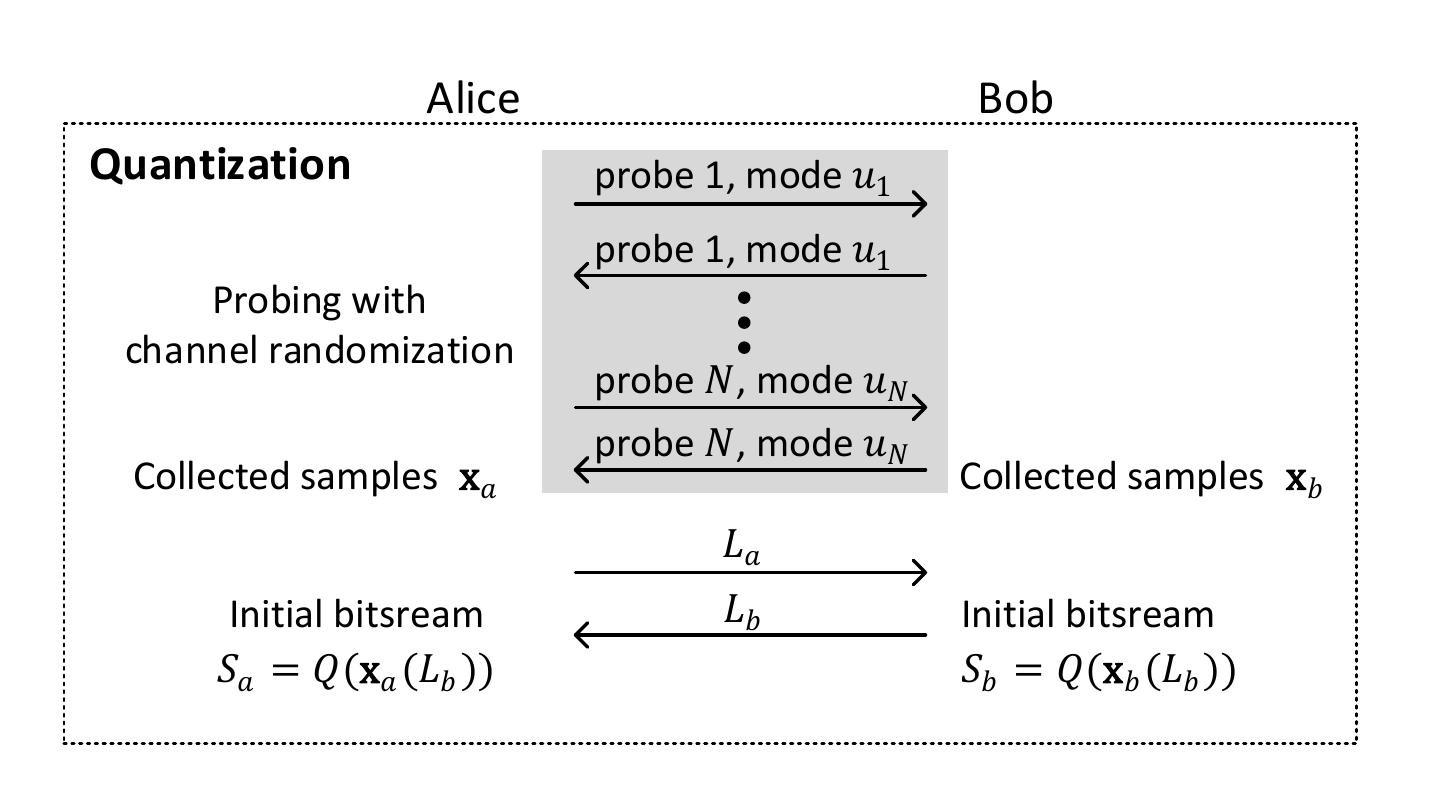}
\end{center}
\vspace{-15pt}
\caption{Quantizition in RAKG: the antenna mode is randomized at every probing round.}
\label{fig:diagram2}
\vspace{-15pt}
\end{figure}

\subsection{Protocol Details}
The key focus of the RAKG protocol is to ensure that the initial bitstreams independently extracted by Alice and Bob during the quantization phase are not predictable by Mallory. Once this security property is established, the security of the entire protocol follows as  the information reconciliation and key verification phases are not influenced by the MitM attack. Hence, we focus our attention on securing the quantization phase. The novel quantization phase  of RAKG is described in the following steps, which are also shown in Fig. \ref{fig:diagram2}. The information reconciliation and key verification phases are the same as the ones shown in Fig. \ref{fig:diagram1}.

\subsubsection{The Quantization Phase in RAKG}
There are two sub-phases:

\noindent\textbf{Probing with channel randomization:}

1.  At the beginning of each probing round, Alice randomly selects an antenna mode and sends a probe to Bob.

2. Bob measures the channel and responds to Alice with his own probe.

3.  Alice measures the channel.
 
4. Alice and Bob repeat steps 1-3 for $N$ probing rounds. At the end of the $N$ rounds, each has a vector of channel measurements $\mathbf{x}_a = \{x_a(1), x_a(2), \dots, x_a(N) \}$ and $\mathbf{x}_b = \{x_b(1), x_b(2), \dots, x_b(N) \}$,  respectively.
 \medskip

\noindent\textbf{Initial bitstream construction:}

5.  Alice and Bob independently select thresholds $q_+^a$, $q_-^a$, $q_+^b$, and $q_-^b$ according to Eq.  \eqref{eq:q}.

6.  Alice compares her channel measurements $\mathbf{x}_a$  with thresholds and sends Bob the list of indexes $L_a$ for which the channel measurements are above $q^a_+$ or below $q^a_-$.

7.  For each index in $L_a$, Bob checks his corresponding measurements in $\mathbf{x}_b$, and makes a list $L_b$ of all indexes with measurements that agree with those of Alice. Bob sends $L_b$ to Alice.

8. Alice and Bob quantize $\mathbf{x}_a$ and $\mathbf{x}_b$  at each index in $L_b$, using the quantization function in Eq.~\eqref{quant} and generate initial bitstreams $S_a$ and $S_b$, respectively.

\medskip



Note that in RAKG, we set the excursion length $e$ equal to one. This is because channel randomization on a per round basis   drastically reduces the excursion length. A short excursion length has been shown to increase the number of extracted bits, but also increasing the probability of bit mismatches, thus establishing two competing factors in the rate of secret key extraction \cite{mathur2008radio}. However, it is beneficial to our protocol in terms of security.

\subsubsection{An Illustrative Example}

\begin{figure}[t]
\vspace{-0.05in}
\centering
\begin{subfigure}[t]{0.52\linewidth}
  \includegraphics[width=\linewidth]{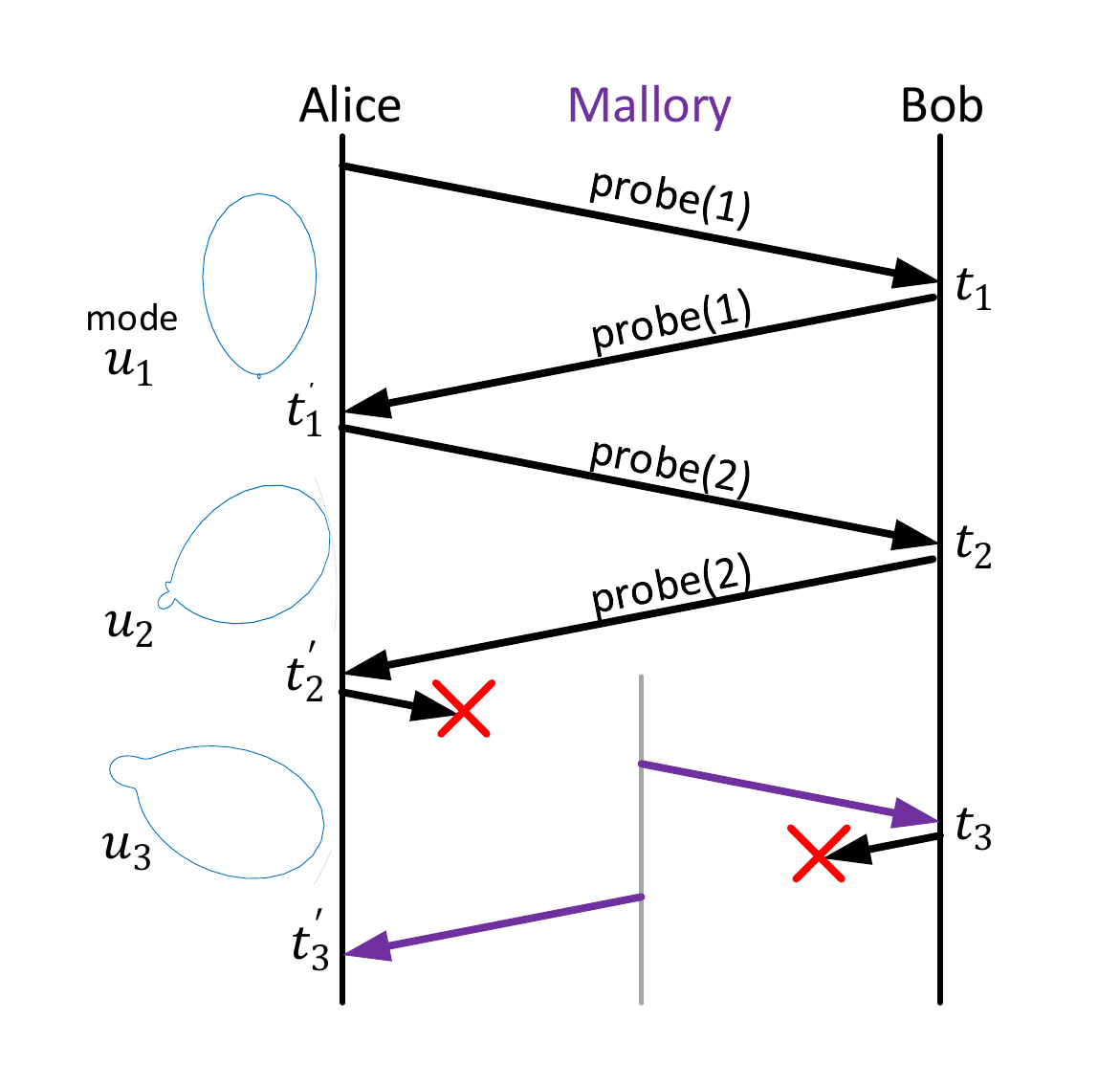}
  \caption{}
\end{subfigure}
\begin{subfigure}[t]{0.47\linewidth}
  \includegraphics[width=\linewidth]{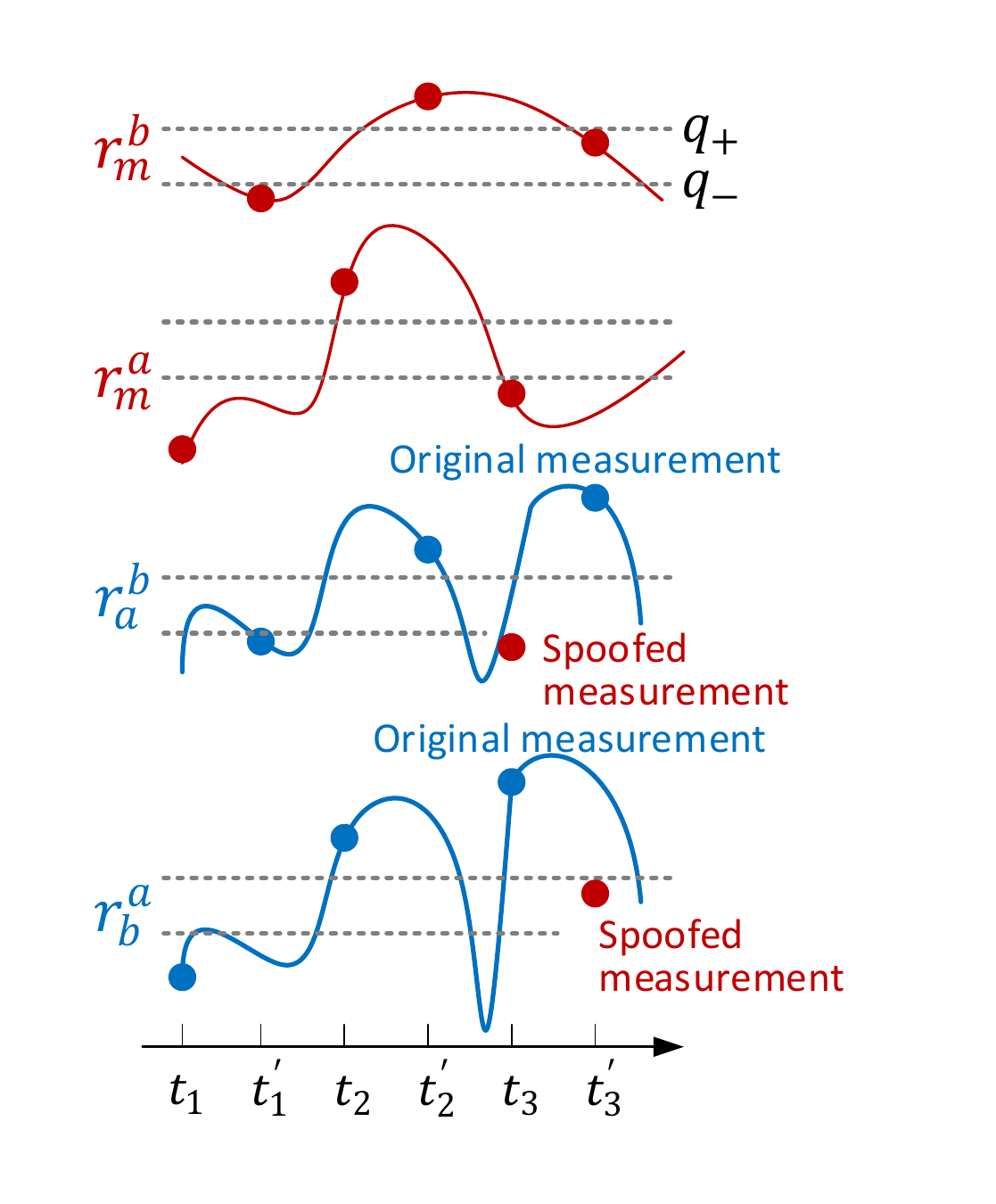}
  \caption{}
\end{subfigure}
\vspace{-5pt}
\caption{(a) Example timeline of the probing phase in RAKG and the MitM attack; (b)  the corresponding channel randomization effect and how the MitM attack is defeated.}
\label{fig:timeline}
\vspace{-15pt}
\end{figure}

Figure \ref{fig:timeline} shows an example timeline   of the channel randomization in RAKG and how it can defend against a MitM attack. For clarity, we denote the time that Bob and Alice receive a probe during  the $i^{th}$ probing round as $t_i$ and $t'_i$, respectively.
In the first probing round, Alice selects antenna mode $u_1$, and no attack opportunity is found by Mallory. In round two, Alice randomly and uniformly selects another antenna mode $u_2$. Mallory indentifies an attack opportunity because $ \text{RSS}_m^a(t_2) \approx \text{RSS}_m^b(t'_2)$ and $\text{RSS}_m^a( t_2) > q_+$. In the third probing round, Alice changes the antenna mode to $u_3$ and exchanges probes with Bob. Mallory jams the probes and injects her own probes to Bob and Alice. The  channel measured by Bob at $t_3$ becomes $\text{RSS}_b^m(t_3)$. Similarly, Alice measures $\text{RSS}_
a^m(t'_3)$. Mallory expects $\text{RSS}_b^m(t_3) \approx \text{RSS}_a^m(t'_3) > q_+$ if the channels stayed  coherent across rounds two and three  (i.e., $\text{RSS}_
m^a(t_3) \approx \text{RSS}_
m^a(t_2)$ and $\text{RSS}_
a^m(t'_3) \approx \text{RSS}_
a^m(t'_2)$).   However, with channel randomization, we  have $h_{am}(t_2)\not\approx h_{ma}(t'_3)$, thus $\text{RSS}_m^a(t'_3) \not\approx \text{RSS}_m^a(t'_2)$ w.h.p. as 
$u_3 \not= u_2$. This introduces a high degree of uncertainty in the bit value guessed by Mallory. 

\section{Security Analysis}
In this section, we evaluate the security of RAKG using theoretical analysis and simulation results. In the analysis, we use the term "key" to refer to the initial bitstreams $S_a$ and $S_b$, rather than considering the keys after the information reconciliation phase. 

\subsection{Security Metrics } \label{sec:metrics}
Consider that after $N$ probing rounds, a key with length $\ell$ is constructed. Let $n$ denote the number of bits that are attacked by Mallory (identified opportunities) of which $m\leq n$ are generated with spoofed measurements. Note that the indices of the attacked bits are known to Mallory whereas the indices of the successes are not. We use the following two metrics proposed in \cite{eberz2012practical} to quantify the success of MitM attack.

    
    
\medskip
\noindent {\bf Key recovery efficiency ($KRE$):} Let $m$ be the  number of bits correctly guessed by Mallory when she finds $n$ opportunities during key generation. The key recovery efficiency is given by
    \[
    KRE = \frac{m}{n}.
    \]
    
\noindent{\bf Key recovery rate ($KRR$):} Let $m$ be the  number of bits correctly guessed by Mallory when a key of length $\ell$ is extracted by Alice and Bob. The key recovery rate is given by
    \[
    KRR =\frac{m}{\ell}.
    \]

Moreover, we define the probability that Mallory correctly guesses the entire key as follows:
\medskip

\noindent\textbf{Key guessing probability} ($p_{\text{key}}$): Let $S'_a$ be the key guessed by Mallory when a key $S_a$ of length $\ell$ is extracted by Alice and Bob. The key guessing probability is given by
    \[
    p_{\text{key}} = \mathbb{P}(S'_a=S_a).
    \]


\subsection{Theoretical Analysis}
To analyze the three metrics defined in Sec.~\ref{sec:metrics},  we first state  Mallory's bit guessing behavior.

{\bf Bit guessing behavior:}
As mentioned in Sec. \ref{sec:MitM}, Mallory's bit guessing behavior is decided by $\text{RSS}_m^a$ in the observed opportunity. 
We define \textbf{opportunity 0} ($O_0$) and \textbf{opportunity 1} ($O_1$) as:
\begin{align*}
    & \textbf{opportunity 0}~(O_0):  \notag \\
    &~\left( |\text{RSS}_m^a - \text{RSS}_m^b| < d \right) ~\textbf{AND}~ \left(\text{RSS}_m^a < q_-, \text{RSS}_m^b < q_- \right) & \\
    & \textbf{opportunity 1}~(O_1): \notag \\
    &~\left( |\text{RSS}_m^a - \text{RSS}_m^b| < d \right) ~\textbf{AND}~ \left(\text{RSS}_m^a > q_+, \text{RSS}_m^b > q_+ \right).
\end{align*}
If Mallory finds an $O_0$, she guesses the bit injected into the next probing round as 0. Similarly, Mallory guesses the injected bit as 1 when $O_1$ is observed. Here we define the opportunity with both $\text{RSS}_m^a$ and $\text{RSS}_m^b$ because they must have the same relationship with $q_+$ or $q_-$ to generate a bit. Hence, we evaluate the attack of Mallory more precisely by considering them jointly.

Based on $O_0$ and $O_1$, here we differentiate the key recovery efficiency for bit 0 and bit 1 at a round $i$ as $p_0$ and $p_1$, which is captured by the following conditional probabilities, respectively.  
\begin{align}
    p_0 & =
    \mathbb{P}\left( \text{RSS}_a^m(i) <q_-, \text{RSS}_b^m(i) <q_- ~\middle|~ O_0 \text{~at~} i-1\right), \label{eq:p0} \\
    p_1 & =
    \mathbb{P}\left( \text{RSS}_a^m(i) >q_+, \text{RSS}_b^m(i) >q_+ ~\middle|~ O_1 \text{~at~} i-1\right). \label{eq:p1}
\end{align}

\begin{proposition}
When antenna modes are independently and uniformly selected for each probing round, $p_0$ and $p_1$ only depend on the state of  the M-A channel during round $i$. That is, $p_0$ and $p_1$ can be simplified as:
\begin{align}
    & p_0  = \mathbb{P}\left( \text{RSS}_a^m(i) <q_-\right), & p_1  = \mathbb{P}\left( \text{RSS}_a^m(i) >q_+\right).
\end{align}
\label{theo1}
\vspace{-10pt}
\end{proposition}
The proof can be found in Appendix, Sec. 8.2.
\medskip

Intuitively, Proposition~\ref{theo1} states that when the channel is randomized with independently and uniformly selected antenna modes, the channel states are independent from round to round. Using the RA multipath channel model in Eq. \eqref{eq:h_u_t}, we derive the closed-form for $p_0$ and $p_1$ as follows:

\begin{proposition} \label{prop2}
Let the M-A channel follow the Rician model with fading parameters  $\nu_{ma}(u)$ and $\varsigma_{ma}(u)$ under antenna mode $u$. If Alice selects antenna modes independently and uniformly at each probing round, and opportunities are found upon RSS, $p_0$ and $p_1$ can be simplified to:
\begin{align}
    p_0 & = \frac{1}{|U|}\sum_{u \in U}(1-Q_1(\frac{\nu_{ma}(u)}{\varsigma_{ma}(u)}, \frac{\sqrt{ 10^{\frac{q_- - P_x}{20}}}}{\varsigma_{ma}(u)})), \label{eq:p_0}\\
    p_1 & = \frac{1}{|U|}\sum_{u \in U}Q_1(\frac{\nu_{ma}(u)}{\varsigma_{ma}(u)}, \frac{\sqrt{ 10^{\frac{q_+ - P_x}{20}}}}{\varsigma_{ma}(u)}). 
    \label{eq:p_1}
\end{align}
\noindent where $Q_1(\cdot)$ is the Marcum Q-function, and $|U|$ is the cardinality of the set of  available antenna modes. 
\label{prop1}
\end{proposition}

The proof can be found in Appendix, Sec. 8.3. 
\medskip

Here, we consider Rician fading since we focus on indoor environments,  but the results can be extended to other channel models such as Rayleigh fading.



Using the individual probabilities $p_0$ and $p_1$, we can now evaluate the probability mass function (PMF) of the number of bits guessed by Mallory using combinatorial arguments. This is derived in the following proposition.

\begin{proposition}
For a key of length $\ell$ extracted by Alice and Bob with RAKG, let  $n$ be the number of opportunities identified by Mallory, and $n_0$ be the number of opportunities for injecting bit 0. The PMF of correctly guessing $m$ injected bits and the average key recovery efficiency and key recovery rates are given by:
\begin{align}
    & \mathbb{P}(M=m) = \sum_{i=0}^m {n \choose m}{m \choose i}p_0^i (1-p_0)^{n_0-i} p_1^{m-i} (1-p_1)^{n-n_0-(m-i)}\\
    & E[KRE]  = \frac{n_0p_0+(n-n_0)p_1}{n},~~~ E[KRR]  = \frac{n_0p_0+(n-n_0)p_1}{\ell}.
\end{align}
\end{proposition}

The proof is a direct application of the binomial distribution assuming independence for each opportunity due to channel randomization.  For recovering the entire key, Mallory guesses the bits being attacked with the bit guessing strategy having a success probability of $p_0$ and $p_1$, respectively. For the remaining bits that do not present an attack opportunity, Mallory can take a random guess. Then the key guessing probability is
\begin{align}
    p_{\text{key}} = 0.5^{\ell-n}p_0^{n_0}p_1^{n-n_0}.
\end{align}

In practice, Mallory may not need to correctly guess the entire key. By exploiting the error correction capacity of the information reconciliation phase,  Mallory can afford up to $w$ errors, where $w$ is the error correcting capability of the code used for reconciliation.

\begin{figure}[t]
\vspace{-0.1in}
\centering
\begin{subfigure}[t]{0.8\linewidth}
  \includegraphics[width=\linewidth]{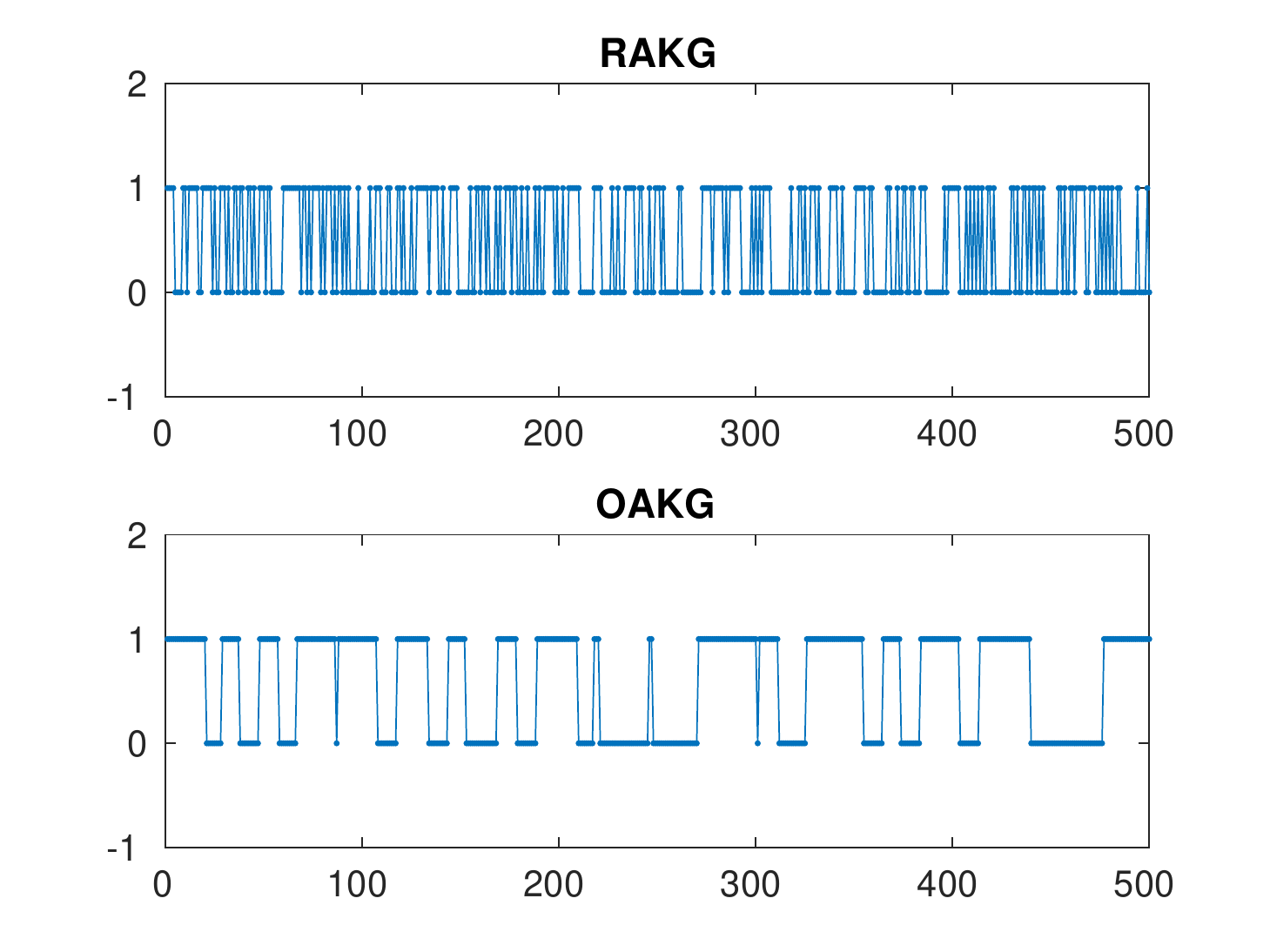}
\end{subfigure}
\vspace{-15pt}
\caption{Example of  the key generated by Alice using RAKG and OAKG,  by assuming the channel coherence time lasts for 10 probing rounds}
\label{fig:ini_key}
\vspace{-10pt}
\end{figure}

\begin{table}
\caption{$p$-value of NIST test results}
\vspace{-0.1in}
\centering
\scalebox{0.8}{
\begin{tabular}{l|cc}
    \toprule
\multirow{2}{*}{NIST Test}
    & &\hspace{-1cm} $p$ value \\
    \cmidrule{2-3}
    & OAKG & RAKG  \\
    \midrule
\multirow{1}{*}{Block Frequency}
    & -- & 0.3505 \\
\multirow{1}{*}{ Runs}
    & -- & 0.3505 \\  
\multirow{1}{*}{ Longest Run}
    & -- & 0.2133 \\   
\multirow{1}{*}{ Rank}
    & 0.7399 &  0.9114 \\
\multirow{1}{*}{ FFT}
    & -- &  0.3505 \\    
\multirow{1}{*}{ Overlapping Template}
    & -- &  0.5341 \\ 
\multirow{1}{*}{ Universal}
    & -- &  0.7399 \\    
\multirow{1}{*}{ Approximate Entropy}
    & -- &  0.5341  \\    
\multirow{1}{*}{ Serial}
    & -- &  0.5341  \\    
    \bottomrule
\end{tabular}}
\label{table:NIST}
\vspace{-5pt}
\end{table}

\begin{table}[h]
\caption{Simulation results of the MitM attack, assuming that the channel coherence time lasts for 10 probing rounds}
\vspace{-0.1in}
	\centering
	\scalebox{0.65}{
	\begin{tabular}{ | c | c | c | c | c |}	
		\hline
		    &     \multicolumn{2}{c|}{RAKG}     &     \multicolumn{2}{c|}{OAKG}\\
		\hline
		$d$    &     2     &     3      &     2     &     3\\
		\hline
        Opportunities[\%]          &    6.66      &     9.57       &     11.97     &     16.00 \\
		\hline
		Bits recovered (0/1)      &          273419/828591     &     399882/1194619     &    1218012/3694757     &     1705445/4776689 \\
		\hline
		Resulting key length     &         36200668      &     36168113     &     33099095     &    33317026 \\
		\hline
        $\mathbf{KRE}$ [\%]    
		&        41.78      &     42.21     &      100     &     100\\
		\hline
		$\mathbf{KRR}$ [\%]     &     3.04     &     4.41      &     14.84     &    19.46 \\
		\hline
	\end{tabular}
	}	
	\label{table:rst1}
	\vspace{-5pt}
\end{table}

\subsection{Evaluation of Security Metrics}
In this section, we evaluate the RAKG security via numerical examples and simulation.  Specifically, we placed Alice, Bob, and Mallory  at  coordinates  $A(5,0)$, $B(15,0)$, and $M(10,5\sqrt3)$, respectively, within an area of $20 \times 10$. Each multipath channel consisted of one LoS path and two NLoS paths produced by two randomly placed scatters within the setup area.
We used the profile data of a customized RA that was designed for channel randomization purpose in \cite{20ROBinYanjun}. The RA has 360 antenna modes, achieved by rotating a log-periodic antenna by one degree. 
One might wonder this RA requires enormous amount of power for the antenna mode that does not point to the A-B channel. We set the signal detection threshold to -75dBm and calculated the minimum transmission power to reach that threshold for each RA antenna mode. We then fixed the transmission power to the maximum power over all modes. Compared with OA, only a 7dBm increase in the transmission power was needed when using the RA mode. This is because the log-periodic RA antenna has higher antenna gains compared with an OA and its 60-degree beamwidth covers a wide range. If devices are further away and power requirements increase, using a subset of more efficient modes would help manage the power requirement.

We first computed $p_0$ and $p_1$ from the RA profile according to Eqs. \eqref{eq:p_0} and \eqref{eq:p_1}, by sampling one million channel states under each antenna mode. 
Eberz et al. \cite{eberz2012practical} proposed an algorithm to estimate $q_+$ and $q_-$ for their MitM attack, where Mallory manipulates a certain number of packets without taking thresholds into account and estimates thresholds according to her attack trace. Their approach yields accurate approximations of $q_+$ and $q_-$. For all of our evaluations, we assume that Mallory knows $q_+$ and $q_-$ set by Alice and Bob for simplicity.
This resulted in $p_0 = 0.53$ and $p_1 = 0.32$. Then, we evaluated the performance of our RAKG protocol by implementing the key generation process and the MitM attack. For comparison, we also implemented our protocol when Alice was equipped with an OA. We call this protocol as OAKG. We set the number of probing rounds to 50 million and the $\beta$ parameter for quantization to 0.4. We assumed that the channel coherence time lasts for ten probing rounds. 

First, we compared the randomness of the generated key with OAKG and RAKG,
to show the impact of our RAKG protocol on increasing channel randomness. In Fig. \ref{fig:ini_key}, We show the first 500 bits of the initial bitstream generated by Alice.  The generated key indicates significant randomness with our RAKG protocol. Further, we tested the key randomness using NIST test suite \cite{bassham2010sp} with the $10^7$-bit length bitstream. The test results are typically in the form of a $p$-value, which must exceed 0.01 for a pass.  We summarized the test results in Table \ref{table:NIST}, where the $p$-values  are all significantly greater than 0.01 for RAKG, while only one of them is beyond 0.01 for OAKG.

We further present the performance of RAKG on defending against the MitM attack and compare it with the OAKG scheme in Table \ref{table:rst1}. We can see that the performance of the adversary with RAKG was much worse than that with OAKG. Noticeably, the $KRE$ and $KRR$ of RAKG is about $1/2$ and less than $1/3$ of OAKG's, respectively. Besides, we observed that the number of recovered 0-bits is less than that of 1-bits in both cases, since there were more $O_1$ than $O_0$ in the current simulation scenario. Generally, opportunities do not occur equally and also $p_0$ and $p_1$ are not equal but vary with  channels and topology, leading to a difference in recovered 0s and 1s. In current simulation, $p_0 = 0.65$ and $p_1 = 0.37$.  Theoretical and simulation results are close, the minor difference between them is likely caused by the  correlation among different antenna modes, since we assumed independent radiation patterns for randomly chosen modes in our security analysis.

Moreover, we computed the average $KRE$ and $KRR$ with the value of opportunities and key length from Table \ref{table:rst1}. We focused on $d=3$, where the adversary earned relatively more advantages with RAKG. The results showed that $E[KRE] = 35.16\%$, $E[KRR] = 3.49\%$ which were smaller than that in Table \ref{table:rst1}. This might because in theory we assume that the M-A, M-B and A-B channels are independent from each other, while they were still some correlations between them, which provide the adversary a better performance than the average case. However, even it was a better case, the $KRE$ and $KRR$ were relatively small. Besides, we compared Mallory's key guessing probability using MitM attack ($p_{\text{key}}$) with random guess. To make $p_{\text{key}}$ greater than the probability of random guess, we should have $\frac{(0.5)^{\ell-
n}\cdot p_0^{n_0} \cdot p_1^{n-n_0}}{0.5^\ell} > 1$, which means $n_0/(n-n_0) > (\ln{0.5}-\ln{p_1})/(\ln{p_0}-\ln{0.5})$. However, 
$n_0/(n-n_0) \approx 0.04$ in Table \ref{table:rst1} while $(\ln{0.5}-\ln{p_1})/(\ln{p_0}-\ln{0.5}) \approx 0.38$, which indicated that the MitM attack was much worse than a random guess when RAKG was implemented. Hence, RAKG is an effective approach to prevent the MitM attack from  recovering portion of the generated key and guessing the entire key.

\section{Performance Evaluation}
\label{sec:eval}

\subsection{Experiment Setup}
\label{ssec:setup}

\begin{figure}[t]
\vspace{-0.05in}
\begin{center}
\includegraphics[width=0.4\textwidth]{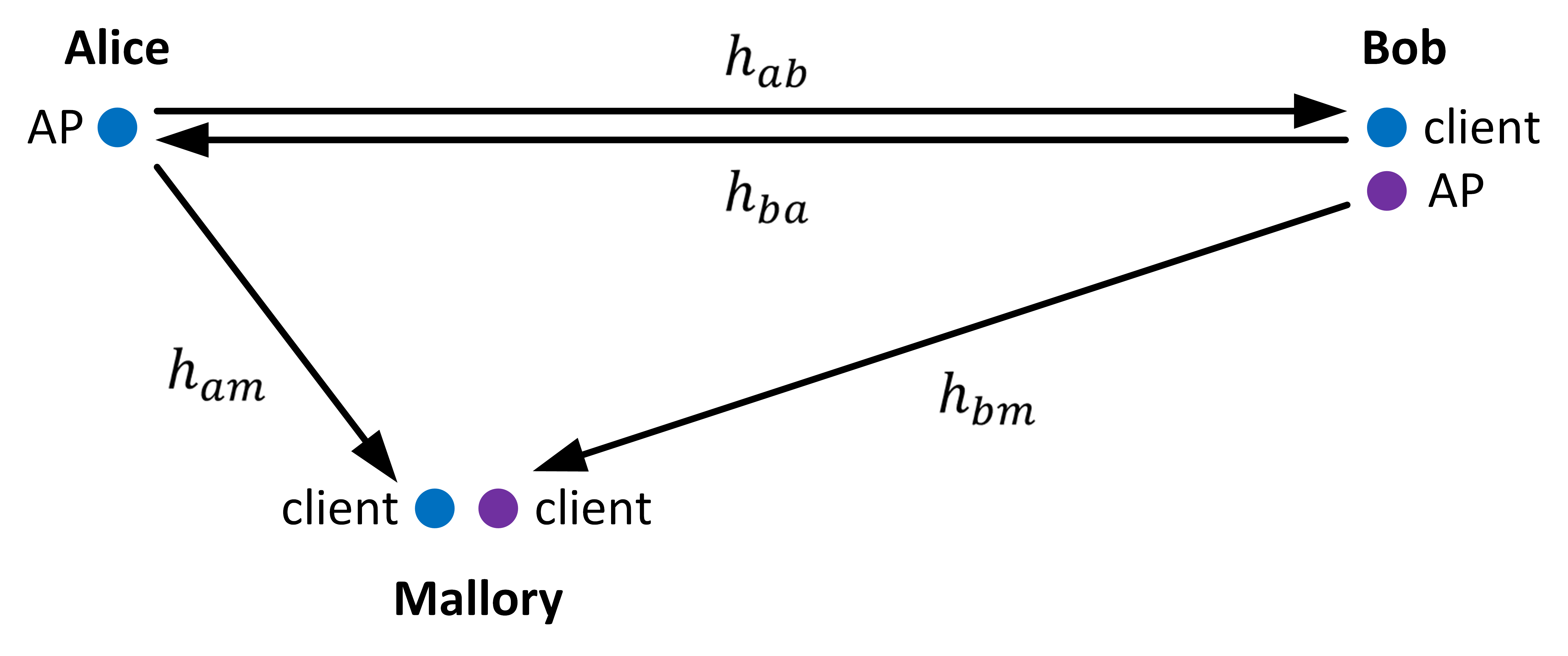}
\end{center}
\vspace{-0.15in}
\caption{The setting of five routers in our experiments with the use of each link}
\label{fig:setting}
\vspace{-15pt}
\end{figure}

We implemented RAKG and OAKG using five commodity Wi-Fi routers (TP-link 4300) and the  Atheros CSI tool \cite{xie2018precise}. Alice and Bob operated in broadcast mode to exchange probes, whereas Mallory monitored both the M-A and M-B channels. However, due to the limitation of this CSI tool's working mode, it can only operate under the broadcast mode. Thus, except for the three routers (shown in Fig. \ref{fig:setting} in blue) to measure $h_{ab}$, $h_{ba}$, and $h_{am}$, we added two more routers (shown in Fig. \ref{fig:setting} in purple) at Bob and Mallory's side to measure $h_{bm}$. We did not implement jamming and injection attacks in our experiments but simulated the attack  with collected data, since \cite{wilhelm2011short} already demonstrated a  reactive jamming attack with a success rate above 99.9\%. In this paper, we focus on evaluating the performance of our RAKG protocol  defending against the MitM attack instead of showing the feasibility of the attack.

We set the probe sampling rate to 20Hz, which translates to a   gap  of 50ms between probes. For RAKG, Alice was equipped with an RA of 3$\times$3 square shaped metallic pixels that are connected by 12 p-i-n diode switches with fast reconfigurability. The radiation patterns of some typical modes can be found in \cite{17TDSCYanjun}, and we switched antenna modes among 4096 available antenna modes. 
However, we only have the data of the antenna patterns for 253 modes, which prevent us from obtaining $p_0$ and $p_1$ among all antenna modes.  Hence, for the experiment results, we did not compute $p_{\text{key}}$. 

\subsection{Metrics for Protocol Performance}

Since our main contribution lies in the quantization phase, we use the following three performance metrics to quantify the performance of the generated random bits (the   initial bitstream after reconciliation), mainly from the efficiency and randomness point of view. These are complementary to  the metrics defined  in  Sec. \ref{sec:metrics} for the MitM attack.
\medskip

\noindent {\bf Bit mismatch rate:} the number of mismatched bits between Alice's and Bob's initial bistreams over the number of total bits in Alice's initial bitstream.

\noindent {\bf Approximate entropy:} captures the randomness and unpredictability of time series. It is preferred over  entropy  because it provides a more accurate  measure when the number of samples is limited \cite{pincus1991approximate}.

\noindent {\bf Secret bit rate:} The average number of secret bits extracted per collected sample. It is measured in terms of final output bits produced after fixing the bit mismatches with information reconciliation, and subtracting the bits revealed to the adversary. 

\subsection{Real-world Experiments}

In this section, we report results derived from data collected from real-world experiments under different scenarios, and compare  the performance of RAKG with OAKG. 

\subsubsection{Experiment A: static indoor environment, same room}
We performed our first experiment inside an apartment unit's living room, where the distance of A-B, M-A, and M-B is 3m, 1.4m, and 2.5m, respectively. The living room was almost empty, and there was no human activity during the experiment. Thus, there was minimal interference to the wireless channel. Figure \ref{fig:exA} shows the raw amplitude of CSI collected by Alice, Bob, and Mallory plotted against the index of probing round. As expected, there are not many variations for all the channels with OAKG, while the A-B and M-A channels are much randomized with RAKG. More importantly, as also demonstrated by \cite{jana2009effectiveness} , when the physical channel is stable, the channel reciprocity is low when OAKG was applied (the curves for Alice and Bob do not follow each other). The variations of the static channel are caused by hardware imperfections and thermal effects that are not reciprocal, as a result, the bit mismatch is relatively high (about 0.5, shown in Table \ref{table:rst_perf}) for OAKG. Thus, it is not possible to extract the secret bit at a fast rate. In contrast, for RAKG, the A-B channel can extract secret bits at a rate of about 0.4, as shown in Table \ref{table:rst_perf}. This is because the channel variations caused by antenna mode change dominated the channel noise.

\begin{table}[h]
	\caption{Protocol performance with MitM attack when $d = 3$}
	\vspace{-0.1in}
	\centering
	\scalebox{0.7}{
	\begin{tabular}{ | c | c | c | c | c | c | c |}	
		\hline
		    &     \multicolumn{3}{c|}{RAKG}     &     \multicolumn{3}{c|}{OAKG}\\
		\hline
        Experiment       &     A     &    B      &     C     &     A     &     B     &     C \\
		\hline
		Mismatch fraction      &     0.2117   &     0.0978    &     0.1513     &     0.4717     &    0.0226     &     0.1813 \\
		\hline
		Approximate entropy     &    0.4914   &     0.5336      &    0.5850     &    --     &     0.2117     &    0.5948 \\
		\hline
		Secret bit rate    &    0.3909     &     0.4551      &     0.3275     &     --      &    0.2645     &     0.3622\\
		\hline
	\end{tabular}
	}	
	\label{table:rst_perf}
	\vspace{-0.2in}
\end{table}

\subsubsection{Experiment B: indoor environment with human activity, same room}
\begin{figure}[t]
\begin{subfigure}[t]{0.25\textwidth}
\includegraphics[width=\textwidth]{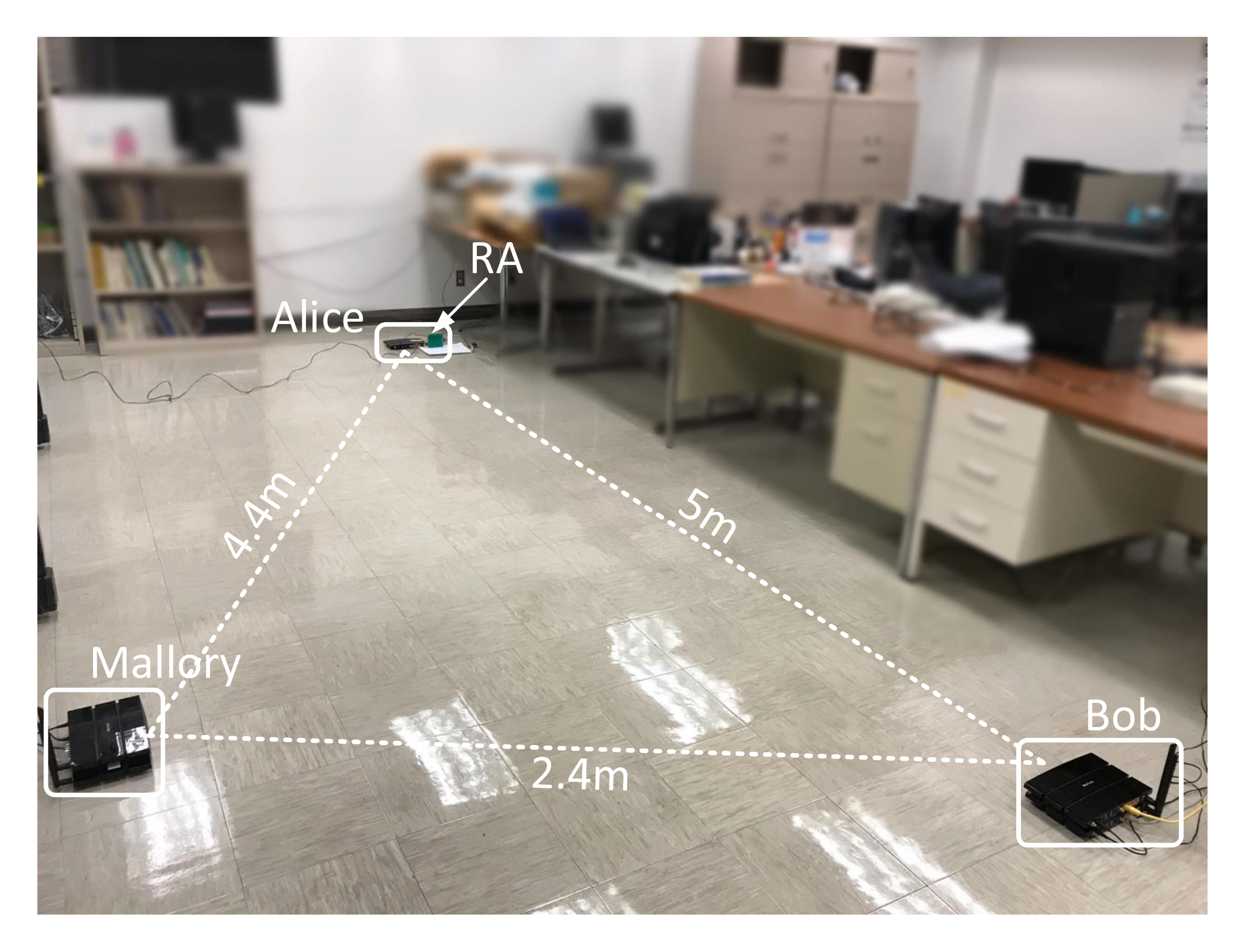}
\caption{}
\label{fig:exB}
\end{subfigure}
\begin{subfigure}[t]{0.2\textwidth}
\includegraphics[width=\textwidth]{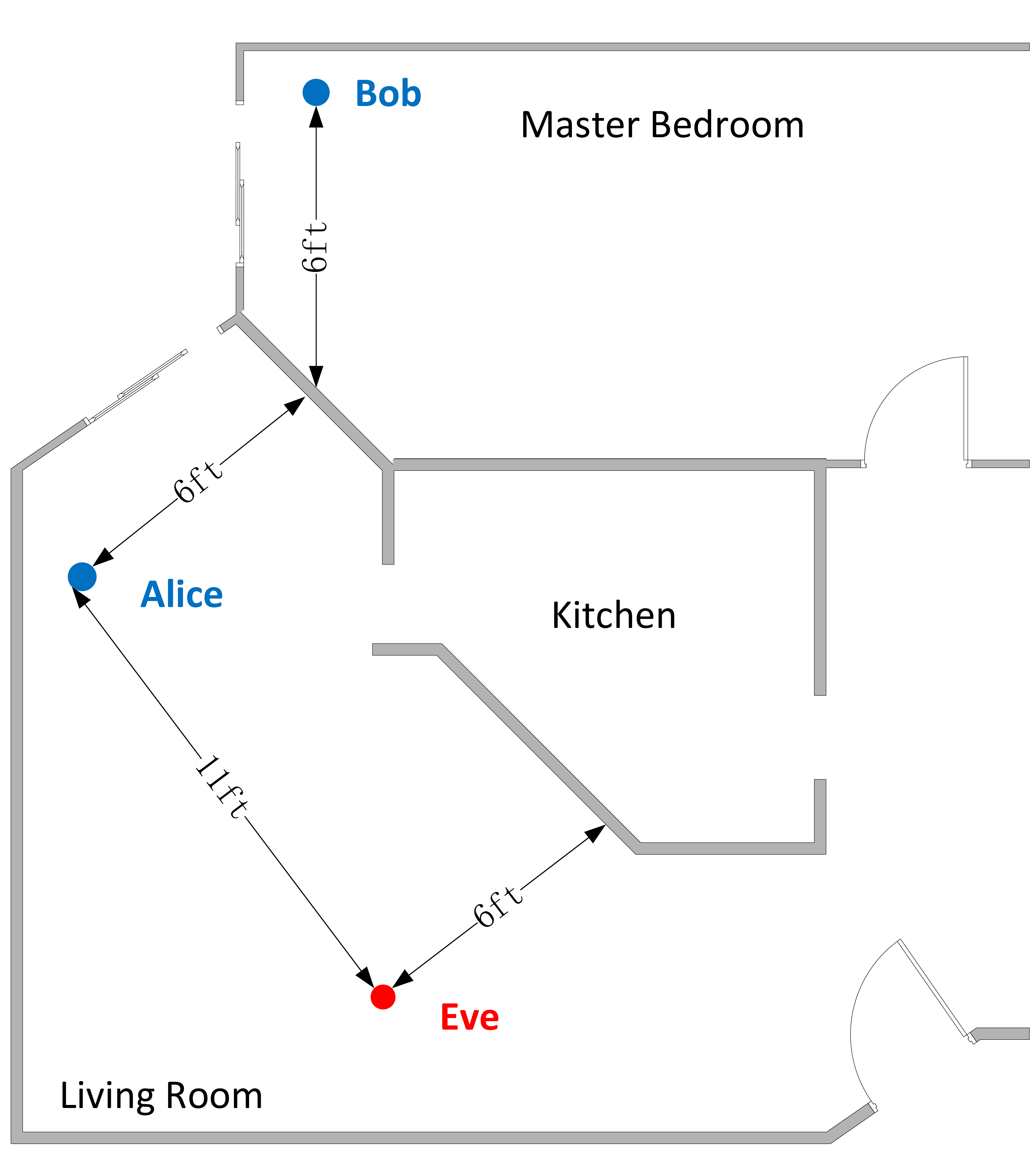}
\caption{}
\label{fig:exC}
\end{subfigure}
\vspace{-0.15in}
\caption{Experiment settings. (a) Experiment B. (b) Experiment C}
\vspace{-15pt}
\end{figure}

\begin{figure*}[t]
\vspace{-0.15in}
\begin{subfigure}[t]{0.48\textwidth}
\includegraphics[width=\textwidth]{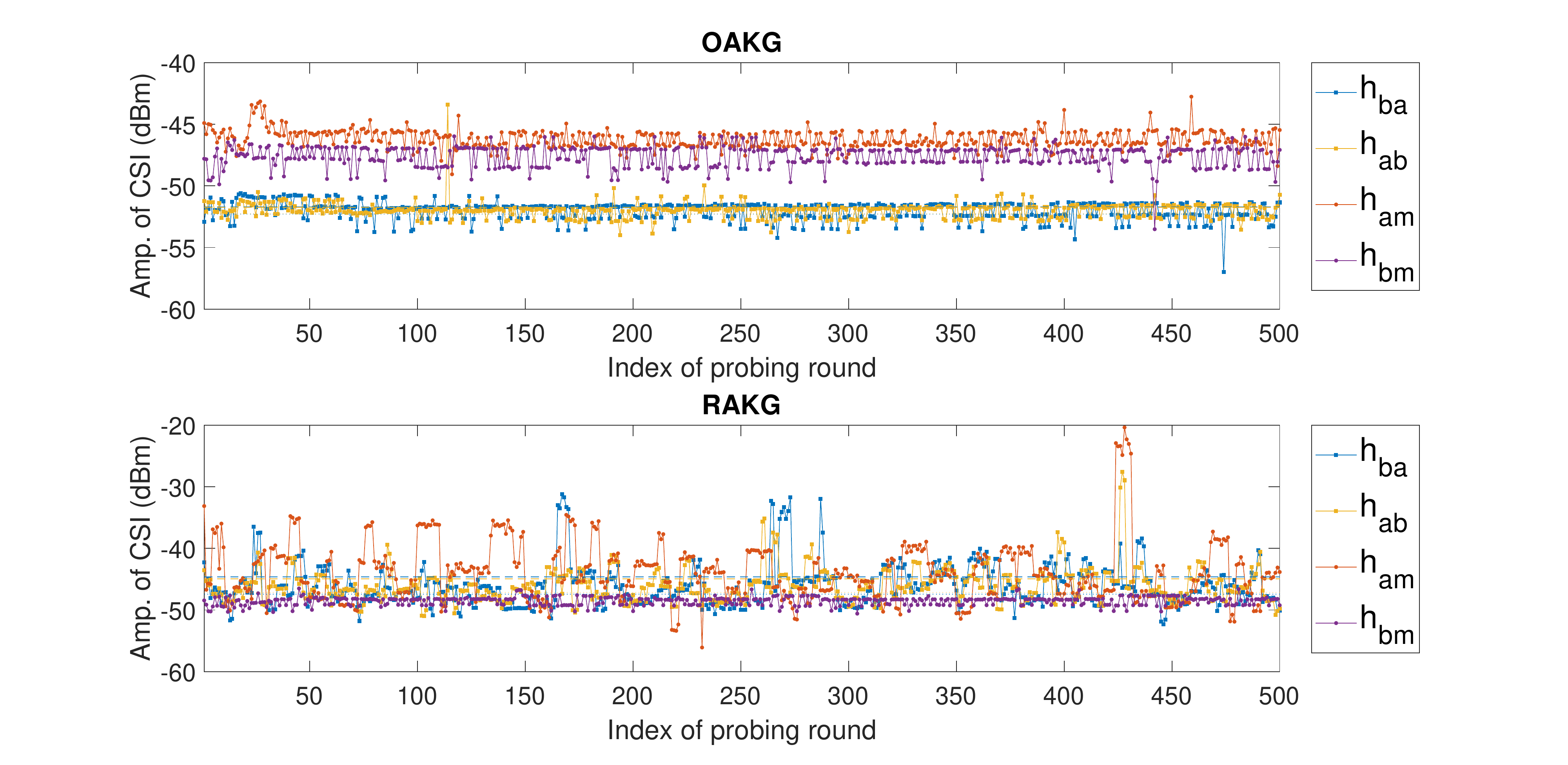}
\caption{}
\label{fig:exA}
\end{subfigure}
\begin{subfigure}[t]{0.48\textwidth}
\includegraphics[width=\textwidth]{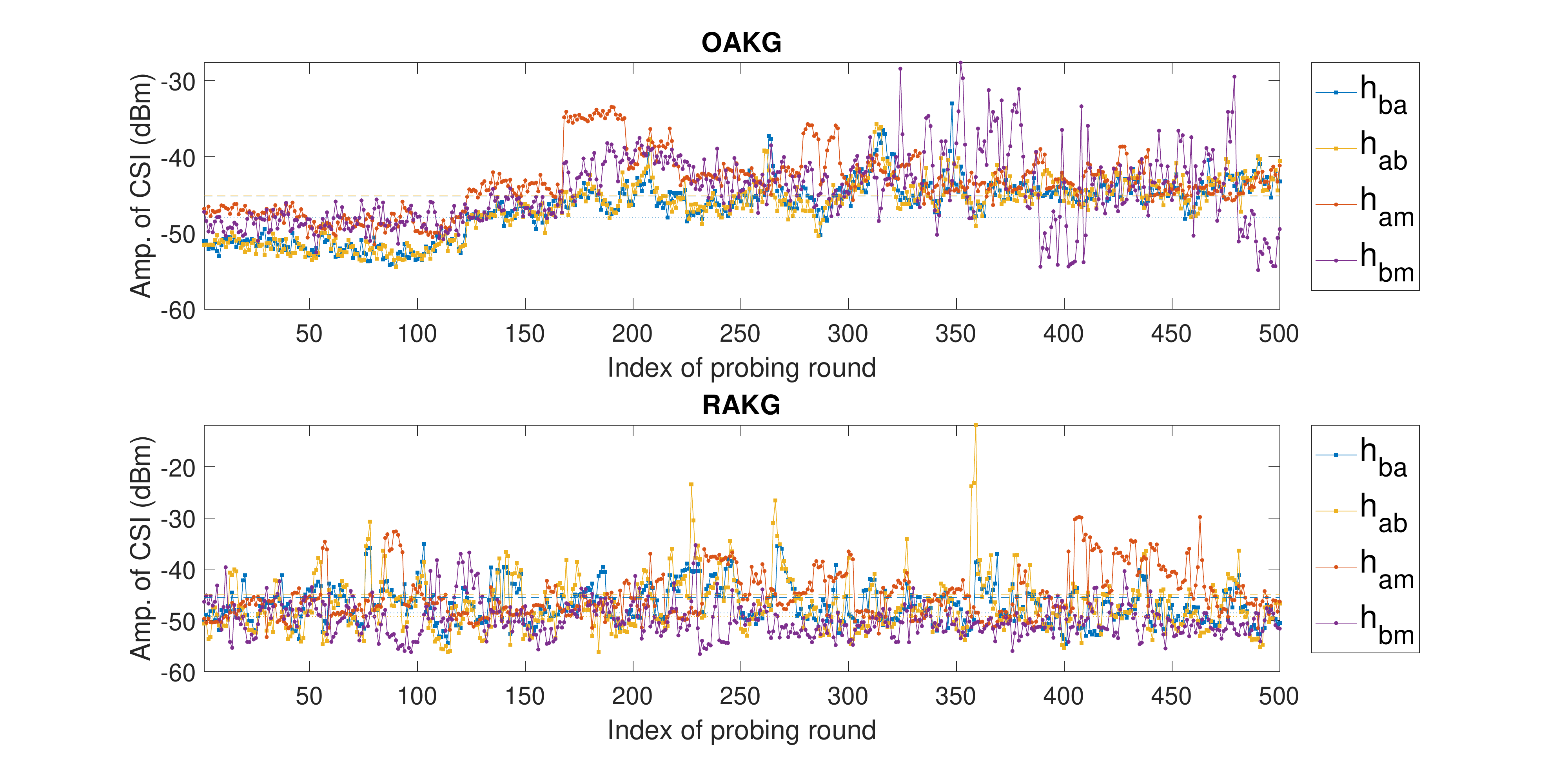}
\caption{}
\label{fig:rstB}
\end{subfigure}
\vspace{-0.15in}
\caption{Amplitude of collected CSI in experiments. (a) Experiment A. (b) Experiment B}
\vspace{-15pt}
\end{figure*}

Next, we did the second experiment in our lab in which involved some human activities. Specifically, we put all the routers on the floor as shown in Fig. \ref{fig:exB}, and let a person move around the area with a vacuum. Figure \ref{fig:rstB} shows the raw amplitudes of CSI collected by Alice, Bob, and Mallory. Compared with the results from Experiment A (Fig. \ref{fig:exA}), we can see that all the channels are randomized, and there is a high degree of reciprocity for the A-B channel. The results of MitM attack are shown in Table \ref{table:rstB}. We can see that under RAKG, both the number of opportunities and key recovery rate 
were reduced. However, we notice that: 1) The revealed bits are  mostly 0-bits under both RAKG and OAKG; this can be explained by the fact that most of the opportunities the adversary found were type $O_0$ in this experiment. However, this is not a real issue if the overall number of bits revealed to Mallory is sufficient, since she is not interested in a specific  key. 2) In this scenario, the result of key recovery efficiency for RAKG is higher than that of the simulation results. This is because the RA we used in the experiment is designed to steer several directions from communication, while we selected the antenna mode in each probing round from all switch combinations at random. The antenna modes with a low gain in the direction of the A-B channel caused packet losses. The channel measurements were likely generated from antenna  radiation patterns with large enough gains. Hence, several adjacent CSI measurements can be similar and bring opportunities to Mallory. Nevertheless, the overall key recovery rate is still low. In conclusion, since  an RA with more distinct antenna patterns was used in the simulation,   the key recovery efficiency of it was much lower than the experiment results. 

\begin{table}[t]
\caption{MitM attack results for Experiment B and C}
\vspace{-0.1in}
	\centering
	\scalebox{0.7}{
	\begin{tabular}{ | c | c | c | c | c | c | c | c| c |}	
		\hline
		\multirow{2}{*}{}& \multicolumn{4}{c|}{Experiment B}& \multicolumn{4}{c|}{Experiment C}\\
		\cline{2-9}
		    &     \multicolumn{2}{c|}{RAKG}     &     \multicolumn{2}{c|}{OAKG}&     \multicolumn{2}{c|}{RAKG}     &     \multicolumn{2}{c|}{OAKG}\\
		\hline
		$d$     &           2     &     3      &         2     &     3&           2     &     3      &         2     &     3\\
		\hline
        Opportunities [\%]      &     6.39      &     10.19      &        18.19     &     23.27 &     7.00      &     8.43      &     23.16     &     25.24 \\
		\hline
		Bits recovered (0/1)      &        11/4     &     15/7     &      89/15     &     114/18 &     8/2     &     12/3     &    41/16     &     43/22\\
		\hline
		Resulting key length     &      316      &     317     &       310     &    310 &        270      &     271     &     281     &    284\\
		\hline
        $\mathbf{KRE}$ [\%]      &     78.95     &     73.33     &    95.41     &     95.65&    58.82      &     65.22     &         69.51     &     70.65\\
		\hline
		$\mathbf{KRR}$ [\%]     &     4.75     &     6.94      &   33.55     &    42.58 &      3.70     &     5.54      &   20.28     &    22.89\\
		\hline
	\end{tabular}
	}	
	\label{table:rstB}
	\vspace{-0.2in}
\end{table}

\subsubsection{Experiment C: indoor environment with human activities,  different rooms}
We performed this experiment in different rooms of an apartment unit and put routers on the desk. The layout   is shown in Fig. \ref{fig:exC}. We involved human activities by letting a person walk in the living room back and forth. The MitM attack results are shown in Table \ref{table:rstB}, which were similar to the results of Experiment B. Both the key recovery rate and key recovery efficiency of OAKG  are reduced significantly compared with Experiment B.  However,   the MitM attack performs better for OAKG if Mallory is closer to  Alice or Bob. In contrast, the distance   has only a small impact on the MitM attack results under RAKG. This is because when the channel is proactively randomized with an RA, the antenna mode has a more significant impact on the multipath gains than human activities. Thus, it makes the channel correlation less sensitive to the distance. 

\section{Related Work}
\noindent {\bf Channel Randomization:} Channel randomization   is viewed as one of the proactive/dynamic defense (or moving target defense) mechanisms. It has been applied for enhancing the security  of wireless communications, in terms of countering both passive and active attacks.




Apart from key generation protocols, for PHY-layer secret communication in general, Haitham et al. \cite{15usenixHaitham}   defend against multi-antenna eavesdroppers in RFID communication, by rotating   eight directional antennas equipped at the transmitter's side using a fan motor and randomly selecting the antennas. Pan et al. proposed  ROBin \cite{20ROBinYanjun} to enhance the security of orthogonal-blinding based secret communication against known-plaintext   attacks by multi-antenna eavesdroppers, also by   varying the channel state  with  rotating directional antennas.


To defend  against signal cancellation attacks (or correlated jamming) and protect  message integrity, 
Pan et al. \cite{17TDSCYanjun}  combine ON-OFF keying modulation with channel randomization via a reconfigurable antenna. The  changing channel prevents the attacker from generating the desired signal to inject, since it would require exact channel states. The signal cancellation attack is quite different from the MitM attack considered in this work, while ON-OFF keying requires PHY-layer   modifications. 

For PHY-layer key generation, it is   challenging to defend against active attacks at the same time. Our work gives a low-cost approach to prevent the MitM attack without using OOB channels, nor requiring changes to the PHY-layer protocol.

\noindent {\bf Physical-Layer Secret Key Generation:}
Since Wyner proposed the  wiretap channel in 1975 \cite{wyner1975wire}, many works advanced the theoretical aspects of physical-layer secret communications \cite{mathur2008radio,wallace2009secure,shiu2011physical}.
Later, Maurer proposed the idea of common randomness, in which two parties
can both tune to a common radio signal source and extract a secret key from it \cite{maurer1993secret}.
Mathur et al.  \cite{mathur2008radio} first proposed a practical RSS and CSI based key generation algorithm for indoor scenarios. Later on, Jane et al. \cite{jana2009effectiveness}  investigated the key generation rate for RSS-based key generation protocols in various channel conditions. It was shown that only highly dynamic mobile scenarios support key generation at a high rate. For static environments, the key generation rate is too low  as the channel lacks randomness. To alleviate this problem,  various approaches have been proposed. For example, Wallace  et al. \cite{wallace2009secure} derived the theoretical key generation performance of the MIMO-based key generation. Zeng \cite{zeng2015physical} designed  an active defensive approach integrating user-generated randomness into pilot signals. Also,  context-based pairing protocols utilize the randomness inherent in
the observed environment to enable devices equipped with the corresponding sensors to extract the key from observed surrounding contents \cite{schurmann2011secure,miettinen2014context}.






Some previous works also used channel randomization  to increase the secret key  rate in key generation protocols.  Specifically, Aono et al. \cite{aono2005wireless} also adopted   reconfigurable antennas to randomize the channel and relied on channel reciprocity to  generate keys. Mehmood et al. \cite{mehmood2014key}  characterized the impact of RA complexity on key generation performance in different static propagation environments.

However, the goal of the above works is to only defend against passive eavesdroppers and increase the secrecy/key rate. Under an active attacker, the adversary can bypass the assumption of channel decorrelation even if it is far  away from the legitimate devices. In particular, Eberz et al. \cite{eberz2012practical} showed the possibility of a MitM attack against   physical-layer key generation protocols \cite{mathur2008radio} by injecting packets to  legitimate transceivers upon observing opportunities, where the attacker can infer up to 47\% of the generated key. Several coutermeasures were suggested by  their work, such as timing-based attack detection, but none of them can effectively prevent MitM attacks.

\noindent {\bf Trust Establishment without Prior Secrets:}
Trust establishment methods aim to establish authenticated secret keys between two or more parties over the wireless channel. Their goal is to  achieve key confidentiality against eavesdroppers, as well as message integrity and authentication   against active attacks. They usually rely on a  cryptographic key exchange protocol  such as Diffie-Hellman, and require out-of-band (OOB) secure channels for authentication \cite{balfanz2002talking,cagalj2006key}. However, OOB channels often require additional hardware interfaces and human interaction, which affects cost, compatibility, and usability. Other works such as I-code \cite{vcapkun2008integrity} proposed in-band approaches to protect message integrity and indirectly authenticate devices through their presence, by exploiting ON-OFF keying and error detection codes. These   works still rely on traditional cryptographic primitives, and thus do not enjoy information-theoretic secrecy for key establishment.

\section{Discussion}

Our adversary model directly follows that proposed in \cite{eberz2012practical}, which does assume the adversary exploits the same properties as legitimate transceivers: the broadcast nature of wireless channels and channel reciprocity.
But in practice, the adversary can have better attack strategies combining channel estimation with power control.
The probe signals sent by Alice and Bob consist of preamble, header, payload, and CRC. Hence, Mallory can estimate the M-A and M-B channels with the preamble. A better attack strategy for Mallory is to jam the legitimate packets once the channel is estimated, and immediately inject packets with different power, to cause similar channel estimates at Alice and Bob. In this way, Mallory does not need to wait for opportunities, she can keep jamming and injection until a significant portion of the generated key is revealed.


This attack is far more complex in terms of required hardware (quick turn around times between receiving, jamming, receiving, etc.) and can be dealt with in three ways. First, we can set the round duration to the duration of exactly one frame exchange so that jamming and transmitting a new frame in each direction would require twice the round duration time. Second, we can reduce the round duration to the bare minimum for channel estimation plus RSS measurement (preamble + some extra symbols) so that Mallory cannot quickly mount an attack without fast switching hardware or multiple receive/transmit chains. Third, we can detect and prevent this attack in the first place. From the perspective of Alice (or Bob), she will receive two consecutive incoming packets in a short time period. The first one cannot be decoded since it is a legitimate packet jammed by Mallory, while the second injected packet   can be decoded. In this case, both Alice and Bob can know that the MitM attack is launched, so that they discard the corresponding estimate. In this way, the MitM attack becomes trivial. The attack implemented by Mallory can only deny the key generation process between Alice and Bob, but never get any knowledge about the generated key.

\section{Conclusions}
In this paper, we proposed a physical-layer key generation protocol which is resistant to  MitM attacks via channel randomization. We leveraged an RA at one of the legitimate transceivers to proactively randomize the channel state across different channel probing rounds. We theoretically analyzed the key guessing capability of the MitM adversary and showed that our approach can be significantly reduce using channel randomization. We conducted extensive simulations and real-world experiments to evaluate the performance of channel randomization. Results show that our RAKG protocol can successfully prevent  the MitM attack by reducing its advantage to nearly random guess. 

\section*{Acknowledgements}
We thank the anonymous reviewers for their helpful comments. This work was partially supported by ARO grant  W911-NF-1910050 and NSF  CNS-1731164.


\section*{Appendix}

\subsection{Key Agreement Scheme (Sch\"urmann
and Sigg \cite{schurmann2011secure})} \label{sec:appendix_reconci}
Denote the initial bitstream generated by Alice and Bob as $S_a$ and $S_b$, respectively. Then it is very likely that $S_a \not= S_b$ because the channel is measured at different times at Alice and Bob (probes are sequentially transmitted) and also due to hardware differences. The information reconciliation approach in \cite{schurmann2011secure} that we adopt (and shown in Fig. \ref{fig:diagram1}) is a fuzzy commitment scheme  utilizing Reed-Solomon (RS) codes to detect and correct the mismatches between $S_a$ and $S_b$. For a RS$(n,k)$ code with $m$-bit symbols, the encoder maps a word of length $k$ uniquely to a specific codeword of length $n$, and the decoder can correct any $t = \lfloor \frac{n-k}{2} \rfloor$ bits
between codewords. We denote the set of 
possible words 
and codewords of RS$(n,k)$ as $\mathcal{Y}$ and $\mathcal{C}$ respectively. 

As an initial step, Alice randomly chooses a 
private word $y \in \mathcal{Y}$
and gets its codeword $c = \text{enc}(y)$, $c \in \mathcal{C}$ with RS encoder. Then a commitment/opening value $\delta$ that does not reveal any information about the 
private word $y$ 
is computed and transmitted to Bob, by $\delta = K_a \ominus c$, where $\ominus$ is the subtract function in $\mathcal{C}$. With $S_b$, Bob calculates a codeword $c'$ with $c' = K_b \ominus \delta$, and recovers the private word with RS decoder, by $y' = \text{dec}(c')$. According to the fuzzy commitment scheme, $y = y'$ only if $\text{Ham}(S_a,S_b) \leq t$, where $\text{Ham}(\cdot)$ is the Hamming distance. Hence, the initial bitstream $S_a$ and $S_b$ can differ in at most $t$ bits, otherwise, the private word cannot be correctly recovered. Finally, Bob can further recover $S_a$ with the add function in $\mathcal{C}$ (i.e., $S_a = \text{enc}(y') \oplus \delta$) and take it as the  key, or Alice and Bob can derive a key from the key derivation function (KDF) with their private words. 

\subsection{Proof of Proposition \ref{theo1}} \label{sec:appendix_1}
\begin{proof}
For simplicity, we analyze $p_1$ and $p_0$ with the CSI amplitude instead of the RSS. 
To proof, we need to start with the multipath channel model  that incorporates the impact
of an RA. In Eq. \eqref{eq:h_u_t}, we described the value of the  time-varing channel at time $t_i$ with a given antenna mode $u_j$. In general, that time-varing  channel can be expressed as the following random variable:
\begin{align}
    h(U, A(t)) = \sum\limits_{l=0}^{P}G(U,\theta_l)A_l(t)
  \label{eq:h1}
\end{align}
where $A_l(t)$ denotes the $l$-th time-varying multipath component and $\theta_l$ is the angle-of-depature of it, $G(U,\theta_l)$ is the random variable that captures the distribution of antenna gain at $\theta_l$ over all the antenna modes. We need to clarify that, the antenna gain $G(U,\Theta)$ is a random variable with some distribution since $U$ can be randomly selected. But for a given antenna mode $u_j$ and departure angle $\theta_l$ of the $l$-th path, $g(u_j,\theta_l)$ is a realization of $G(U,\Theta)$ that is fixed, which means $G(U=u_j,\Theta = \theta_l) = g(u_j,\theta_l)$. Hence, $h(U,A(t))$ in Eq. \eqref{eq:h1} is a time-varying random variable that also depends on the antenna mode. We use the specific antenna mode $u_j$ and time instance $t_i$ to denote a realization of it, that is, $h(U = u, A(t) = a(t_i)) = h(u,a(t_i))$, and to simplify, we rewrite $h(u_j,a(t_i))$ as $h(u_j,t_i)$, and gets Eq. \eqref{eq:h_u_t}
\begin{align*}
    h(u_j, t_i) & = \sum\limits_{l=0}^{P}g(u_j,\theta_l)a_l(t_i)
\end{align*}
where $a_l(t_i)$ is the realization of $A_l(t)$ at time $t_i$, that is, $A_l(t = t_i) = a_l(t_i)$.

For clarity, we denote the time that Bob and Alice receive a probe during  the $i^{th}$ probing round as $t_i$ and $t'_i$, respectively. Denote the antenna mode used at round $i$ as $u_i$.
Assume that $O_1$ is found at probe round $i-1$, then Mallory implements the attack at $t_i$ and guesses the injected bit as 1, then we can express $p_1$ as:
\begin{align}
    p_1 & = \mathbb{P}\left( \overline{h}_{ma}(U_i,A_{ma}(t'_i)) > q_+, \overline{h}_{mb}(A_{mb}(t_i)) > q_+ ~\middle|~ \right. \notag \\
    & \qquad\quad |\overline{h}_{am}(U_{i-1},A_{am}(t_{i-1})) - \overline{h}_{bm}(A_{bm}(t'_{i-1}))| < d, \notag \\
    & \qquad\quad \left. \overline{h}_{am}(U_{i-1},A_{am}(t_{i-1})) > q_+, \overline{h}_{bm}(A_{bm}(t'_{i-1})) > q_+ \right)
    \label{eq:p1_1}
\end{align}
where $\overline{h}_{xy}$ denotes the amplitude of the  $x$-$y$ channel. To consider the worst case for the protocol, we assume the physical channel remains static from finding an opportunity until finishing the attack, which means:
\begin{align}
    A_{xy}(t_{i-1}) = A_{xy}(t'_{i-1}) = A_{xy}(t_i) = A_{xy}(t'_i), ~\forall x\text{-}y \text{~channel} 
\end{align}
Besides, with channel reciprocity, we have
\begin{align}
    A_{yx}(t_i) & = A_{xy}(t_i) ~\forall x\text{-}y \text{~channel}
\end{align}

Then we ignore the index of time in CSI and use $A_{xy}$ to denote the time-varing multipath components, Eq. \eqref{eq:p1_1} can be rewritten as:
\begin{align}
    p_1 & = \mathbb{P}\left( \overline{h}_{ma}(U_i,A_{ma}) > q_+, \overline{h}_{mb}(A_{mb}) > q_+ ~\middle|~ \right. \notag \\
    & \qquad\qquad |\overline{h}_{am}(U_{i-1},A_{am}) - \overline{h}_{bm}(A_{bm})| < d, \notag \\
    & \qquad\qquad \left. \overline{h}_{am}(U_{i-1},A_{am}) > q_+, \overline{h}_{bm}(A_{bm}) > q_+ \right) \label{eq:p1_21}\\ 
    & = \mathbb{P}\left( \overline{h}_{am}(U_i,A_{am}) > q_+ ~\middle|~ \right. \notag \\
    & \qquad\qquad |\overline{h}_{am}(U_{i-1},A_{am}) - \overline{h}_{bm}(A_{bm})| < d, \notag \\
    & \qquad\qquad \left. \overline{h}_{am}(U_{i-1},A_{am}) > q_+, \overline{h}_{bm}(A_{bm}) > q_+ \right)
    \label{eq:p1_2}
\end{align}
where we simplify Eq. \eqref{eq:p1_21} with channel coherence and reciprocity. 

Obviously, for the channel $h(U,A)$,  the antenna mode part $U$ and the multipath part $A$ are independent. Hence, Eq. \eqref{eq:p1_2} can be expressed as:
\begin{align}
    p_1 & = \int_{-\infty}^{+\infty}\mathbb{P}\left( \overline{h}_{am}(U_i,a_{am}) > q_+  ~\middle|~\right. \notag \\ 
    & \quad|\overline{h}_{am}(U_{i-1},a_{am}) - \overline{h}_{bm}(A_{bm})| < d_{\max}, \notag \\
    &\quad \left. \overline{h}_{am}(U_{i-1},a_{am}) > q_+, \overline{h}_{bm}(A_{bm}) > q_+ \right)f_{A_{am}}(a_{am})da_{am} \label{eq:p1_3}
\end{align}
by using the property that the joint probability of two independent random variables is the product of their marginal probabilities, where $f_{A_{am}}(a_{am})$ denotes the distribution of $A_{am}$ that includes all the time-varying  multipath components of the M-A channel.

As the antenna modes for different probing rounds are independently and uniformly selected, then $h_{am}(U_{i-1},a_{am})$ and $h_{am}(U_i,a_{am})$ are independent. Besides, the M-A, M-B, and A-B channels are independent from each other since Mallory is more than half-wavelength away from both Alice and
Bob. Therefore, the right-hand side and left-hand side of Eq. \eqref{eq:p1_3} are independent, and it can be simplified as:
\begin{align}
    p_1 & = \int_{-\infty}^{+\infty}\mathbb{P}\left( \overline{h}_{am}(U_i,a_{am}) > q_+\right) f_{A_{am}}(a_{am}) d a_{am} \\
    & = \mathbb{P}\left( \overline{h}_{am}(U_i,A_{am}) > q_+\right) \\
    & = \mathbb{P}\left( \overline{h}_{ma}(U_i,A_{am}) > q_+\right)
    \label{eq:hh2}
\end{align}
with channel reciprocity.

Also, we have:
\begin{align}
    p_0 & = \mathbb{P}\left( \overline{h}_{ma}(U_i,A_{ma}(t_i)) < q_-, \overline{h}_{mb}(A_{mb}(t_i))< q_- ~\middle|~ \right. \notag \\
    & \qquad\qquad |\overline{h}_{am}(U_{i-1},A_{am}(t_{i-1}) - \overline{h}_{bm}(A_{bm}(t_{i-1}))| < d, \notag \\
    & \qquad\qquad \left. \overline{h}_{am}(U_i,A_{am}(t_{i-1})) < q_-, \overline{h}_{bm}(A_{bm}(t_{i-1})) < q_- \right) \\
    & = \mathbb{P}\left( \overline{h}_{ma}(U_i,A_{am}) < q_-\right)
\end{align}

Similarly, when RSS is considered, we have 
\begin{align*}
    p_0 & =
    \mathbb{P}\left( \text{RSS}_a^m(i) <q_-, \text{RSS}_b^m(i) <q_- ~\middle|~ O_0 \text{~at~} i-1\right) \\
    & = \mathbb{P}\left( \text{RSS}_a^m(i) <q_-\right) \\
    p_1 & =
    \mathbb{P}\left( \text{RSS}_a^m(i) >q_+, \text{RSS}_b^m(i) >q_+ ~\middle|~ O_1 \text{~at~} i-1\right) \\
    & = \mathbb{P}\left( \text{RSS}_a^m(i) >q_+\right)
\end{align*}
\end{proof}

\subsection{Proof of Proposition \ref{prop2}}
To proof, we start with the distribution of the CSI amplitude. Then we simplify $p_0$ and $p_1$ based on CSI amplitude. Finally, we show the simplified $p_0$ and $p_1$ when opportunities are found upon RSS.

{\bf Distribution of the CSI amplitude:} According to Eq. \eqref{eq:h1}, the multipath channel under any antenna mode $u$ is:
\begin{align}
    h(u, A(t)) = \sum\limits_{l=0}^{P}g(u,\theta_l)A_l(t)
    \label{eq:hh1}
\end{align}
where each time-varying multipath component can be expressed as $A_l(t)= X_l(t) + jY_l(t)$, where $X_l(t)$ and $Y_l(t)$ are independent and identically distributed (iid) stationary time-varying Gaussian random variables with the same variance $\sigma_0^2$, and mean $\mu_{x,l}$, $\mu_{y,l}$, respectively. 

For indoor scenario, denote $l = 0$ as the LoS component, and $l = 1,\cdots,P$ as the NLoS components, we have:
\begin{align}
& \mu_{x,l} =\left\{
    \begin{aligned}
        \mu_{x,0}, &\quad l = 0 \\
        0, &\quad l = 1,\cdots, P\\
    \end{aligned}
\right. 
& \mu_{y,l} =\left\{
    \begin{aligned}
        \mu_{y,0}, &\quad l = 0 \\
        0, &\quad l = 1,\cdots, P\\
    \end{aligned}
\right.
\end{align}
Then Eq. \eqref{eq:hh1} can be rewritten as:
\begin{align}
    h(u, A(t)) & = \sum\limits_{l=0}^{P}g(u,\theta_l)(X_l(t) + jY_l(t)) \notag \\
   & = \left( g(u,\theta_0)X_0(t) + \cdots +g(u,\theta_P)X_P(t) \right) + \notag \\
   &\qquad\quad j\left( g(u,\theta_0)Y_0(t) + \cdots + g(u,\theta_P)Y_P(t) \right) \notag \\
   & = X(u,t) + jY(u,t)
\end{align}
where $X(u,t)$ and $Y(u,t)$ are iid stationary time-varying Gaussian random variables with the same variance $\sigma^2(u)$. The mean of them is $\mu_x(u)$ and $\mu_y(u)$,  respectively.
\begin{align}
    & \sigma^2(u) = \sum\limits_{l=0}^{P} \left( g(u,\theta_l)\sigma_0 \right)^2 = \sigma^2_0 \sum\limits_{l=0}^{P}  g^2(u,\theta_l)
    \\
    & \mu_x(u) = \sum\limits_{l=0}^{P}g(u,\theta_l)\mu_{x,l} = g(u,\theta_0)\mu_{x,0} \notag \\
    & \mu_y(u) = \sum\limits_{l=0}^{P}g(u,\theta_l)\mu_{y,l} = g(u,\theta_0)\mu_{y,0}
\end{align}
which means for any antenna mode $u$, the amplitude of the multipath channel (i.e., $\overline{h}(u, A(t))$), follows Rician distribution with PDF:
\begin{align}
    \frac{\overline{h}(u, A(t))}{\varsigma^2(u)} e^{-\frac{\overline{h}^2(u, A(t)) + \nu^2(u)}{2\varsigma^2(u)}} I_0\left( \frac{\overline{h}(u, A(t))\nu(u) }{\varsigma^2(u)} \right)
\end{align}
where  $I_0(\cdot)$ is the modified Bessel function of the first kind with order zero, and parameters $\nu(u)$ and $\varsigma(u)$ are
\begin{align}
    & \nu(u) = \sqrt{\mu_x^2(u) + \mu_y^2(u)} = g(u,\theta_0) \sqrt{\mu_{x,0}^2 + \mu_{y,0}^2} \notag \\
    & \varsigma(u) = \sigma_0 \sqrt{\sum\limits_{l=0}^{P}  g^2(u,\theta_l)}
\end{align}
and the mean and the variance of $\overline{h}(u, A(t))$ are
\begin{align}
    & \mu(u)  = \varsigma(u) \sqrt{\pi/2}L_{1/2}\left(-\frac{\nu^2(u)}{2\varsigma^2(u)}\right) \\
    & \sigma^2(u) = 2\varsigma^2(u) + \nu^2(u) - \frac{\pi \varsigma^2(u)}{2}L^2_{1/2}\left(-\frac{\nu^2(u)}{2\varsigma^2(u)}\right)
\end{align}
where $L_{1/2}(\cdot)$ is the Laguerre polynomial, and $L^2_{1/2}(\cdot)$ is the square of it.

{\bf Simplify $p_0$ and $p_1$ when opportunities are found upon CSI amplitude:} With the distribution of $\overline{h}(u, A(t))$ above, Eq. \eqref{eq:hh2} can be further expressed as:
\begin{align}
    p_0 & = \mathbb{P}\left( \overline{h}_{ma}(U_2,A_{ma}) < q_-\right) \notag \\
    & = \sum_{u \in U} \mathbb{P}\left( \overline{h}_{ma}(u,A_{ma}) < q_- \right)p_U(u)\\
    & = \sum_{u \in U} (1-Q_1(\frac{\nu_{ma}(u)}{\sigma_{ma}(u)}, \frac{q_-}{\sigma_{ma}(u)}))p_U(u)\\
    & = \frac{1}{|U|}\sum_{u \in U}(1-Q_1(\frac{\nu_{ma}(u)}{\varsigma_{ma}(u)}, \frac{q_-}{\varsigma_{ma}(u)}))
\end{align}
where $p_U(u)$ is the PMF of antenna mode selection, which is $\frac{1}{|U|}$ by assuming the antenna mode is uniformly selected.

Similarly,
\begin{align}
    p_1 & = \mathbb{P}\left( \overline{h}_{ma}(U_2,A_{ma}) > q_+\right) \notag \\
    & = \frac{1}{|U|}\sum_{u \in U}Q_1(\frac{\nu_{ma}(u)}{\varsigma_{ma}(u)}, \frac{q_+}{\varsigma_{ma}(u)})
\end{align}
Note that, since the opportunities are found upon CSI amplitude, $q_+$ and $q_-$ are found upon CSI amplitude as well.

{\bf Simplify $p_0$ and $p_1$ when opportunities are found upon RSS:}
When Alice transmits with mode $u$, for A-B and M-A channels the received signal can be represented as
\begin{align}
    y(u,t) = h(u,A(t))x(t)
    \label{eq:y}
\end{align}
where $x(t)$ is the transmitted signal with transmit power $\sigma_x^2$. The noise of the channel is ignored in Eq. \eqref{eq:y} for simplicity. Then, the received signal strength (RSS) in dB or dBm is 
\begin{align}
    \text{RSS} & = 20\log(\overline{h}(u,A)) + P_x 
\end{align}
where $P_x$ is the transmit power, and the notation of time in $\overline{h}(u,A(t))$ is ignored for simplicity.\\
Then
\begin{align}
    p_0 & =  \mathbb{P}\left( \text{RSS}_a^{m} < q_- \right) \notag \\
    & = \mathbb{P}\left( 20\log ( \overline{h}_{ma}(u,A))  + P_x < q_- \right)\\
    & = \mathbb{P} \left( \overline{h}_{ma}(u,A) < \sqrt{ 10^{\frac{q_- - P_x}{20}}} \right) \\
    & = \frac{1}{|U|}\sum_{u \in U}(1-Q_1(\frac{\nu_{ma}(u)}{\varsigma_{ma}(u)}, \frac{\sqrt{ 10^{\frac{q_- - P_x}{20}}}}{\varsigma_{ma}(u)}))
\end{align}
Similarly, 
\begin{align}
    p_1 & =  \mathbb{P}\left( \text{RSS}_a^{m} > q_+ \right) \notag \\
    & = \frac{1}{|U|}\sum_{u \in U}Q_1(\frac{\nu_{ma}(u)}{\varsigma_{ma}(u)}, \frac{\sqrt{ 10^{\frac{q_+ - P_x}{20}}}}{\varsigma_{ma}(u)})
\end{align}\\
$q_+$ and $q_-$ here are found upon RSS, as the opportunities and channel measurements are found based on RSS.

\bibliographystyle{ACM-Reference-Format}
\bibliography{refpaper}


\end{document}